\documentclass[preprint,amsmath,amssymb,aps,prd,showpacs]{revtex4-1}
\usepackage{graphicx}
\usepackage{dcolumn}
\usepackage{bm}
\usepackage{color}
\def\m{\mathop{\rm m}\nolimits}

\begin{document}

\title{ Kinetic dominance and psi series in the Hamilton-Jacobi formulation of inflaton models}
\author{Elena Medina}
\email{elena.medina@uca.es} \affiliation{
                Departamento de Matem\'aticas,
                Facultad de Ciencias,
                Universidad de C\'adiz,
                11510 Puerto Real, Spain}
\author{Luis Mart\'{\i}nez Alonso}
\email{luism@ucm.es} \affiliation{
            Departamento de F\'{\i}sica Te\'orica II,
                Facultad de Ciencias F\'{\i}sicas,
                Universidad Complutense,
                28040 Madrid, Spain}

\date{\today}
\begin{abstract}
Single-field inflaton models in the kinetic dominance period admit formal solutions given by generalized asymptotic  expansions called  psi series. We present a method for computing  psi series for the Hubble parameter as a function of the inflaton field in the Hamilton-Jacobi formulation of inflaton models.  Similar psi series  for the scale factor, the conformal time and the Hubble radius are also derived. They are applied to determine the  value of the inflaton field when  the inflation period starts and to estimate  the contribution of the kinetic dominance period to calculate the duration of inflation.  These psi series  are also used  to obtain explicit two-term  truncated psi series near the singularity for  the potentials of the Mukhanov-Sasaki equation for curvature and tensor perturbations. The method is illustrated  with  wide families of inflaton models  determined by potential functions combining  polynomial and exponential functions as well as  with generalized Starobinsky models.

 \end{abstract}
\maketitle

\section{Introduction}\label{intro}

The theory of inflationary cosmology provides a  framework to study the early universe~\cite{STA80,GU81,LI82} which solves several central problems of the hot Big Bang model. In the present work we consider  single-field  inflaton models formulated in terms of a time dependent  real field $\phi(t)$ in a spatially flat universe ~\cite{LI85,MU05,BA09,MA18}. For an homogeneous spatially flat Friedman-Lema\^{i}tre-Robertson-Walker (FLRW) spacetime with scale factor $a(t)$, these models are described by the nonlinear ordinary second order differential equation
\begin{equation}\label{eq1}
\ddot{\phi}+3H\dot{\phi}+\frac{dV}{d\phi}(\phi)\,=\,0,
\end{equation}
where  $H=\dot{a}/a$ is the Hubble parameter which is related to the inflaton field by
\begin{equation}\label{eq2}
H^2=\frac{1}{3m_{Pl}^2}\left(\frac{1}{2}\dot{\phi}^2+V(\phi)\right).
\end{equation}
 Here $V=V(\phi)$ is a given potential function,  $m_{Pl}=\sqrt{\hbar c/8\pi G}$ is the Planck mass and  dots indicate derivatives with respect to the cosmic time $t$.

We concentrate on the \emph{kinetic dominance} (KD) period~\cite{LA05,HAN14,HAN19} of inflaton models,  when the kinetic energy of the inflaton field dominates  over its potential energy.
\begin{equation}\label{kd}
\dot{\phi}^2\gg V(\phi).
\end{equation}
It is a non-inflationary or pre-inflationary stage  that is  followed by a short fast-roll inflation phase ~\cite{DE10} and afterwards by the traditional slow-roll inflation stage \cite{MU05,STEIN84,STE93,LID94,LIDS97,BA06,LI08,WE08,BO09}. Recently,  Handley et al~\cite{HAN14,HAN19,HAD18,HER19,HER19case} have  shown the relevance of the KD period  (\ref{kd})  for setting initial conditions. In fact, as they prove in \cite{HAN14}, under mild conditions on the potential $V$, all solutions (except perhaps a single one) evolve from a KD region. Our study is devoted to the   asymptotic series solutions of the inflaton equations for the KD period and their applications.

The solutions of  the equation~(\ref{eq1}) manifest generically  branch point singularities of logarithmic type.  This is the same type of singularities exhibited by the solutions of the Lorentz system  \cite{TA81} and their presence is associated to  the so-called \emph{ psi series}~\cite{HI97} asymptotic solutions of ordinary differential equations.  Alternative different  psi series containing terms with irrational or even complex exponents were found  in the H\'enon-Heiles system~\cite{CH83}, the rigid body problem, the Toda lattice equation, the Duffing oscillator~\cite{FO88} and the fractional Ginzburg-Landau equation~\cite{TA06}.
Logarithmic psi series solutions of the inflaton equations~(\ref{eq1})   have been also  considered, see for instance ~\cite{LA05,DE10,HAN14}.
 Recently~\cite{HAN19}  a general method has been formulated  for computing psi series expansions  for the solutions of the equations~(\ref{eq1}) and the generalization of~(\ref{eq2})
for FLRW spacetimes with curvature.  The method formulates the inflaton equations as a four-dimensional first-order system of ordinary differential equations  and determines solutions as series expansions involving  powers of $t-t^*$ and $\log(t-t^*)$. These series are termed  \emph{logolinear series} in ~\cite{HAN19}.

In the present  work we  propose an alternative
 method  for determining formal asymptotic solutions of ~(\ref{eq1})-(\ref{eq2}) expressed as psi series. It uses the Hamilton-Jacobi formalism of the inflaton models  \cite{SA90}, \cite{LI00} and \cite{SEP}, in which the independent variable is the inflaton field instead of the cosmic time.
We generate  psi series  for solutions of the Hamilton-Jacobi equations  for the Hubble function as  functions of  the rescaled inflaton field $\varphi \equiv  \sqrt{\frac{3}{2}}\phi/m_{Pl}$. We apply our method to the following  classes of models :
\begin{enumerate}
\item
\begin{equation}\label{pot10}
v(\varphi)\,=\,\sum_{n=0}^Nv_n(\varphi)e^{-n\varphi},
\end{equation}
where $v_n(\varphi)$ are polynomials in $\varphi$ and $N$ is a non-negative integer.

\item Models with generalized Starobinsky  potential functions
\begin{equation}\label{pot20}
v(\varphi)\,=\,\sum_{n=0}^Nv_n\,e^{-\alpha n \varphi},
\end{equation}
where $ \alpha $ is any irrational number and $v_n$ are constant coefficients.
\end{enumerate}
In~(\ref{pot10}) and~(\ref{pot20}) $v(\varphi)$ stands for the rescaled potential $v(\varphi)=3V(\phi)/m_{Pl}^2$.

The paper is organized as follows.  In section~\ref{hj} we briefly
introduce the Hamilton-Jacobi formalism of inflation models.
Section~\ref{psi} describes  our method for determining psi series  for the  inflaton models with potential functions (\ref{pot10}) and (\ref{pot20}).
 For the case (\ref{pot10})   we determine
a one-parameter family of  logarithmic  psi series solutions
in  the variable $u:= e^{-\varphi}$ with polynomial coefficients depending on $\varphi$, which could be termed \emph{expolinear series} . In particular,
the models with polynomial potentials  ($N=0$)  exhibit several interesting
symmetry properties which are analyzed  in detail.  As illustrative
explicit examples we apply the method to the quadratic potential
$v(\varphi)=\m^2\varphi^2$ and to
 the Higgs potential $v(\varphi)=g^2(\varphi^2-\lambda^2)^2$.
For the models with potential functions (\ref{pot20})  we characterize a
one-parameter family of  psi series solutions  of non-logarithmic
type in the variable $u:= e^{-\varphi}$ with  coefficients  which are polynomials in   $e^{-\alpha\varphi}$.  Furthermore, a simple limit operation shows that the results  also apply to rational exponents  $\alpha$.
 In particular, we apply the method to the Starobinsky potential
 $$v(\varphi)\,=\,\lambda\left(1-e^{-\alpha\varphi}\right)^2,$$
and check that our results with $\alpha=\pm\frac{2}{3}$ coincide with the results in~\cite{HAN19} for the corresponding inflaton model with potential
\begin{equation}\label{sll}
V(\phi)=\Lambda^2\Big(1-e^{-\sqrt{\frac{2}{3}}\phi} \Big)^2.
\end{equation}
 At the end of section ~\ref{psi} we discuss how to derive from our psi series in the inflaton field $\varphi$  the logolinear series  involving  powers of $t-t^*$ and $\log(t-t^*)$.

Finally, section~\ref{app} presents several applications of the psi series obtained  in the previous section to calculate analytical approximations of several relevant quantities of inflation models and to compare them with the corresponding numerical approximations. Thus, we use the psi series to determine the value of the inflaton field  at the initial moment of the  inflation period. We also provide a formula for the amount of inflation, which includes  the contribution of the part of the KD period  which overlaps the inflation region.  Finally, we consider the potentials of the Mukanov-Sasaki equation near the singularity for both curvature and tensor perturbations. It is known ~\cite{DE10} that as functions of the conformal time the dominant  term  of these potentials coincides with the critical central singular attractive potential allowing the fall to the center of a quantum particle. Then, we use the psi series previously obtained  to provide an explicit  two-term truncated psi series approximation to these potentials.

\section{HAMILTON-JACOBI FORMULATION OF INFLATON MODELS}\label{hj}

From equations~(\ref{eq1})-(\ref{eq2}) it follows that
\begin{equation}\label{dH}
\dot{H}=-\frac{1}{2m_{Pl}^2}\dot{\phi}^2,
\end{equation}
As a consequence the Hubble parameter  $H$ is a positive monotonically decreasing function of $t$. This property implies that  for smooth and positive  potential functions  $V$,  the solutions $\phi(t)$  of~(\ref{eq1})
with arbitrary finite initial data  do not have singularities forward in the cosmic time $t$.  Nevertheless,  the function $H(t)$ increases without bound  backwards in time,
so that $H(t)$ and $\phi(t)$ may develop  singularities.

The presence of singularities backwards in time can be  expected from
the following argument: If the  KD condition (\ref{kd}) holds
then  we may neglect $V$ and $V_{\phi}$  in the inflaton equations  and  from~(\ref{eq1})  we have
\begin{equation}
    \label{appr}
    \ddot{\phi} +  \sqrt{\frac{3}{2}}\frac{1}{m_{Pl}}|\dot{\phi}  |\, \dot{\phi} \sim 0.
\end{equation}
Thus we obtain two families of  approximate solutions
\begin{equation}\label{phiasy}
\phi\sim\pm\sqrt{\frac{2}{3}}m_{Pl}\log(t-t^*)+\phi_p \quad \mbox{as} \quad t\rightarrow(t^*)^+,
\end{equation}
where $t^*$  and $\phi_p$ are arbitrary constants. The corresponding asymptotic form of the Hubble parameters is
\begin{equation}\label{hasy}
H\sim\frac{1}{3(t-t^*)} \quad \mbox{as} \quad t\rightarrow(t^*)^+.
\end{equation}
These approximate solutions of the inflaton equations are the dominant terms of the psi series expansions that we will consider below.

\subsection{The Hamilton-Jacobi equations}

 We use the rescaled variables
\begin{equation}\label{newvar}
\varphi=\sqrt{\frac{3}{2}}\frac{\phi}{m_{Pl}}, \quad v(\varphi)=\frac{3}{m_{Pl}^2}V(\phi),\quad h=3 H,
\end{equation}
and rewrite  equations ~(\ref{eq1}) and (\ref{eq2}) as
\begin{equation}
    \label{eq:me}
    \ddot{\varphi} + h\dot{\varphi} + \frac{1}{2}v'(\varphi)= 0,
\end{equation}
and
\begin{equation}
    \label{eq:htau}
    h^2= \dot{\varphi}^2+ v(\varphi),
\end{equation}
respectively.

In order to discuss  inflaton models in the Hamilton-Jacobi formalism we consider a reduced  space of initial conditions  $(\varphi,\dot{\varphi})\in \mathbb{R}^2$ for (\ref{eq:me})-(\ref{eq:htau}) such that
\begin{equation}
    \label{dpm}
    \varphi\geq \varphi_0, \quad \dot{\varphi}<0,
\end{equation}
where $\varphi_0$ will be assumed to be a fixed value of $\varphi$ such that  the  potential $v$ and its first derivative $v'$ are smooth
and strictly positive  for $\varphi\geq \varphi_0$ .
The map $(\varphi,\dot{\varphi}) \mapsto (\varphi,h)$
enables us to describe the dynamics of  (\ref{eq:me})-(\ref{eq:htau}) on the  subset
\begin{equation}
    \label{Rpm}
    R=\{(\varphi,h)\in \mathbb{R}^2 : \varphi\geq \varphi_0, \quad \sqrt{v(\varphi)}<h<+\infty\},
\end{equation}
of the $(\varphi,h)$ plane.  The Hamilton-Jacobi formulation of the equations ~(\ref{eq1})-(\ref{eq2}) is given
by the couple of equations
 \begin{equation}\label{e1}
h'(\varphi)^2\,=\,h(\varphi)^2-v(\varphi),
\end{equation}
and
 \begin{equation}\label{e}
\dot{\varphi}=-h'(\varphi),
\end{equation}
Here primes denote derivatives with respect to $\varphi$ and
the Hubble function $h$ is  assumed to be the positive root
\begin{equation}\label{htau2}
    h= \Big(\dot{\varphi}^2+ v(\varphi)\Big)^{1/2}.
\end{equation}
The set $R$ plays the role of the phase space of the formalism.
Each  solution $h=h(\varphi)$ of  (\ref{e1}) determines  a corresponding  implicit solution $\varphi(t)$ of (\ref{e}) given  by
\begin{equation}
    \label{imp}
    t=-\int_{\varphi(0)}^{\varphi(t)}\frac{d \varphi}{h'(\varphi)}
\end{equation}

From the physical point of view, the early universe is assumed to emerge from a state with energy density  $3m_{Pl}^2 H $ of
the same order of the Planck density $m_{Pl}^4$.
Below that density  the classical inflationary description of the universe is not valid. Hence  the only physical constraint required for the
initial data  of a classical inflationary universe  is that its energy density  $3m_{Pl}^2 H $ should not be larger than
$m_{Pl}^4$ or, equivalently, $h\le\,m_{Pl}^2$ which in Planck units ($G=c=\hbar=1)$ means
\begin{equation}\label{cl}
 h< h_{p}:=\frac{1}{8 \pi}\,\approx\,0.0398.
\end{equation}

\subsection{Inflation and kinetic dominance  regions}

The inflation period of the universe evolution is characterized by
an accelerated universe expansion $\ddot{a}>0$. From the identity
\begin{equation}\label{ad2}
\frac{\ddot{a}}{a}=\frac{1}{3m_{Pl}^2}\left(V(\phi)-\dot{\phi}^2\right),
\end{equation}
it follows that this period is determined by the constraint
\begin{equation}\label{in}
\dot{\phi}^2<V(\phi).
\end{equation}
Then
 it follows at once  that the
inflation region (\ref{in}) in $R$  is characterized  by
\begin{equation}\label{in2}
\sqrt{v}<h<\sqrt{\frac{3 v}{2}}.
\end{equation}
For a successful solution to the \emph{horizon} and \emph{flatness} cosmological problems it is required that the  \emph{amount of inflation} during the period of inflation
\begin{equation}\label{ef}
  N=\int_{\phi_{in}}^{\phi_{end}}\frac{H}{\dot{\phi}} d \phi=\frac{1}{3}\int_{\varphi_{end}}^{\varphi_{in}}\frac{h}{h'} d \varphi,
\end{equation}
should be close to $N\sim 60$ \cite{BA09,BA12,DO03,MA18}.
Thus,  given a solution $h=h(\varphi)$ of  (\ref{e1}) it is  important to determine the values $\varphi_{in}$, $\varphi_{end}$  for which inflation starts and ends, respectively. Due to (\ref{in2}) both values  satisfy
\begin{equation}\label{inf}
h(\varphi)=\sqrt{\frac{3 v(\varphi)}{2}}.
\end{equation}

 The approximate solutions (\ref{phiasy}) of (\ref{eq1}) correspond to approximate solutions of (\ref{e1}) of the form
 \begin{equation}\label{hvarphiasy}
h\sim\frac{e^{\pm\varphi}}{b}\quad \mbox{as} \quad \varphi\rightarrow\pm\infty,
\end{equation}
where $b$ is an arbitrary strictly positive parameter.
Due to the symmetry $(v(\varphi),h(\varphi),\varphi(t))\rightarrow (v(-\varphi),h(-\varphi),-\varphi(t))$
of the equations~(\ref{e1})-(\ref{e}),  without loss of generality we restrict our analysis to series expansions of solutions of~(\ref{e1}) such that
\begin{equation}\label{h1}
h\sim \,\frac{e^{\varphi}}{b}\quad \mbox{as} \quad \varphi\rightarrow +\infty.
\end{equation}
Solutions $h=h(\varphi)$ of (\ref{e1})  which have the  asymptotic form (\ref{h1}) emerge  from the KD region  and blow up at a finite time $t=t^*$ given by
\begin{equation}
    \label{impb}
    t^*=-\int_{\varphi(0)}^{\infty}\frac{d \varphi}{h'(\varphi)}.
\end{equation}

\begin{figure}
\begin{center}
                 \includegraphics[width=9cm]{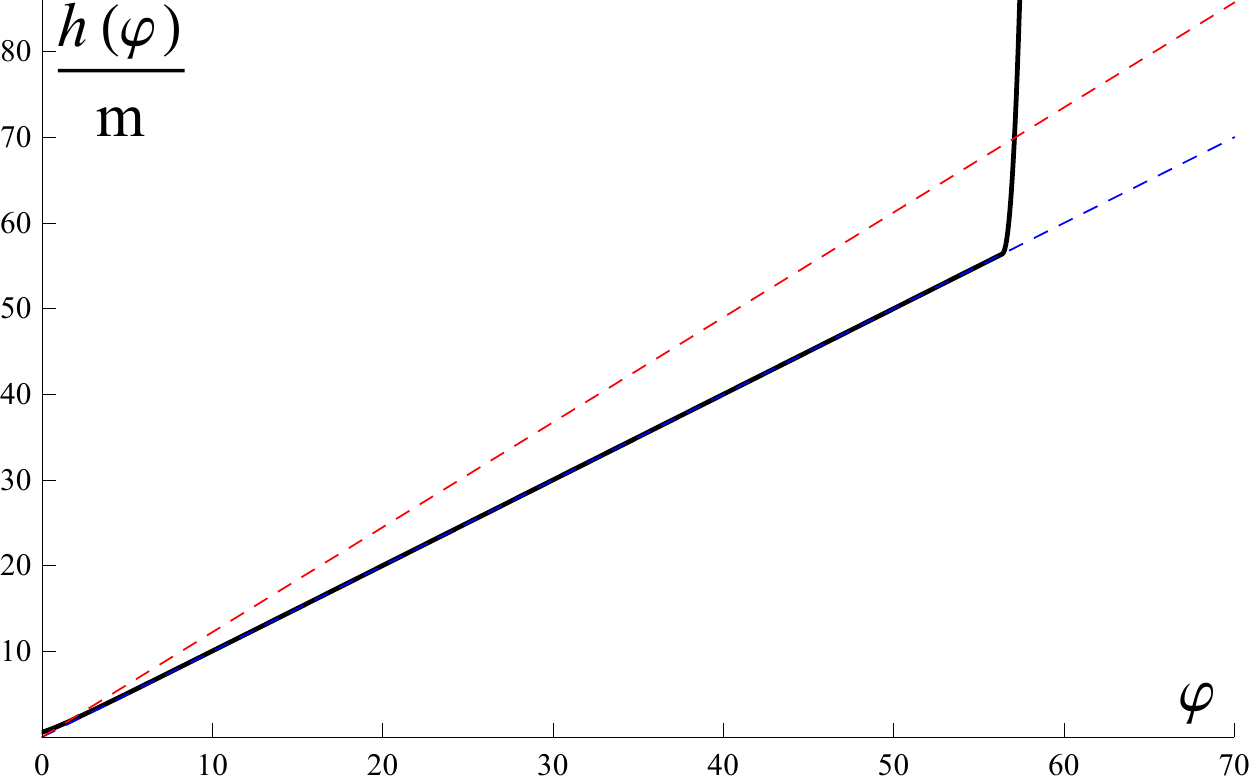}
\end{center}
\caption{Reduced hubble parameter for the quadratic model
$v(\varphi)=\m^2\varphi^2$. The black line shows the rescaled
numerical solution $h/\m=h(\varphi)/\m$  of (\ref{e1}) such that
$h(60)=10^3\m$. The region between blue and red  dotted lines is
the inflation region.
}
\end{figure}
\begin{figure}
\begin{center}
        \includegraphics[width=7cm]{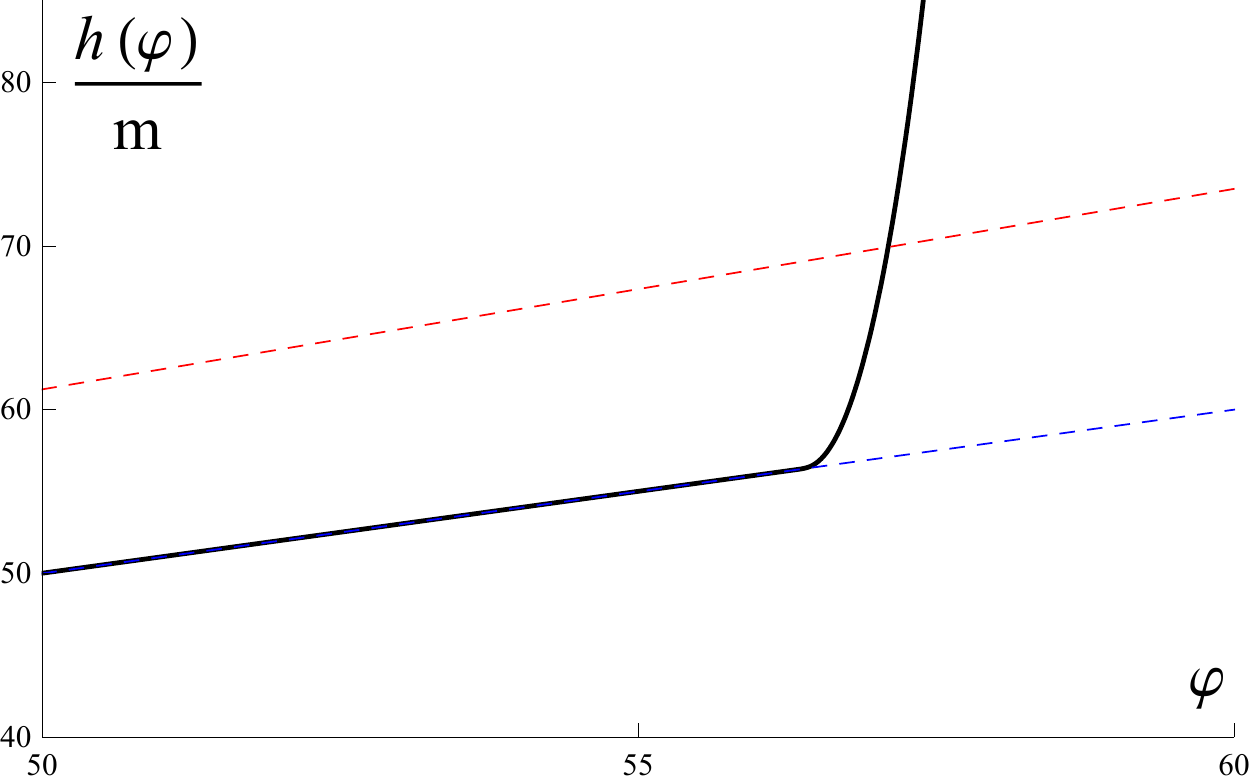}
        \hspace{1cm}
        \includegraphics[width=7cm]{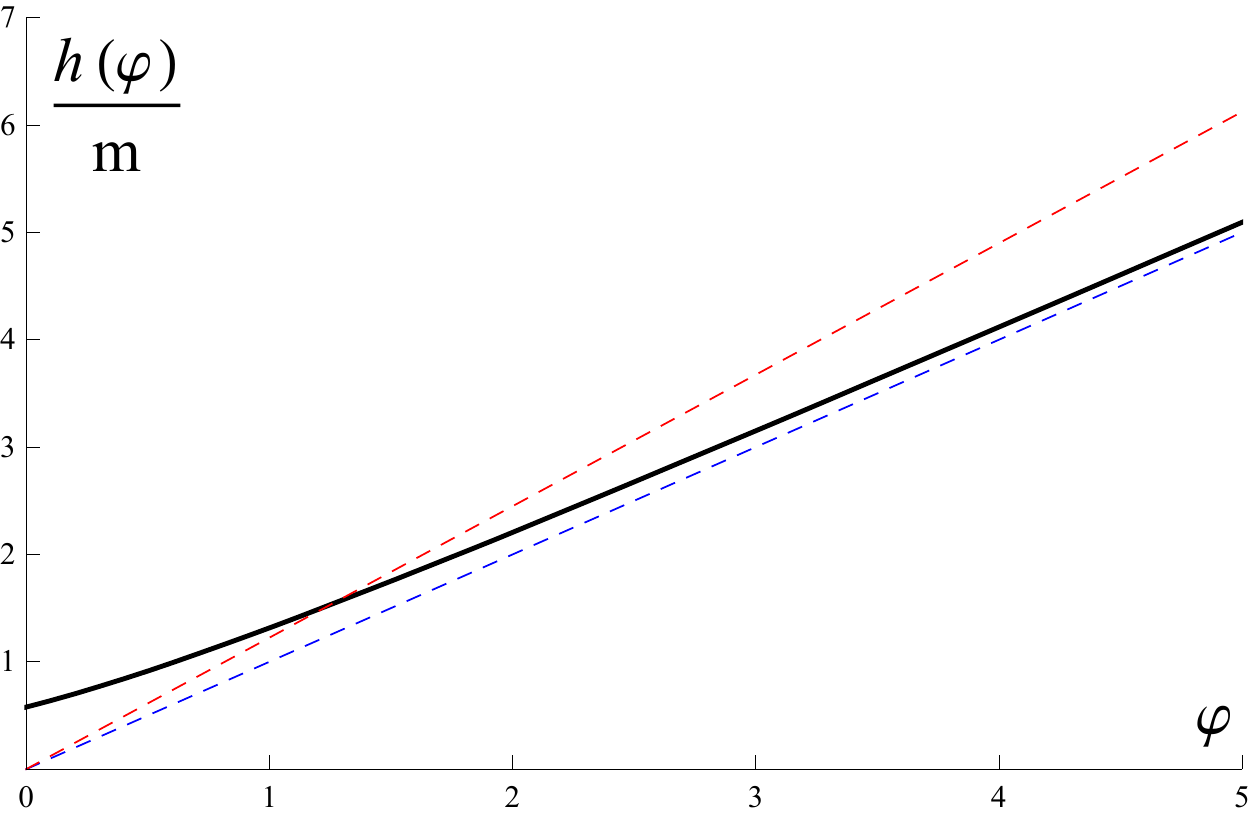}
\end{center}
    \caption{ The left figure shows the solution of Figure 1 entering  the inflation region in a KD regime and
    approaching the slow-roll regime  (blue dotted line). The right figure shows the same solution leaving the
    inflation region.  }
\end{figure}

\pagebreak



\section{PSI SERIES IN THE HAMILTON-JACOBI FORMULATION}\label{psi}

In this section we determine a one-parameter family of  psi series formal solutions of the equation~(\ref{e1})  for the  models with  \emph{polynomial-exponential potentials } (\ref{pot10}) and
Starobinsky potentials (\ref{pot20}).

\subsection{Psi series for polynomial-exponential potentials}

Let us consider the differential equation~(\ref{e1}) for a potential $v(\varphi)$ of the family~(\ref{pot10}).  We look  for  psi series solutions of the form
\begin{equation}\label{hpot100}
h(\varphi)\,=\,\frac{e^{\varphi}}{b}\,+\,\sum_{n=1}^{\infty}h_n(\varphi)e^{-(n-1)\varphi},
\end{equation}
where the  coefficients $h_n(\varphi)$ are polynomial functions of
$\varphi$ and  $b$ is a non-zero positive real parameter.

By substituting~(\ref{hpot100}) into (\ref{e1}),
we obtain
\begin{equation}\label{eq1h}\everymath{\displaystyle}\begin{array}{l}
\frac{2}{b}\sum_{n=1}^{\infty}(h_n'-nh_n)e^{-(n-2)\varphi}\,+\,\sum_{n=0}^Nv_n(\varphi)e^{-n\varphi}\,=\,\\  \\
\qquad -\sum_{n=2}^{\infty}\left[\sum_{j+k=n}(h'_j-(j-1)h_j)(h'_k-(k-1)h_k)-h_jh_k\right]e^{-(n-2)\varphi}.
\end{array}\end{equation}
Then, identifying  the coefficients of $e^{\varphi}$ in~(\ref{eq1h}) we have that $h_1$ must satisfy the equation $h'_1-h_1=0$,  whose  polynomial solution is
\begin{equation}\label{pot1h1}
h_1(\varphi)=0.
\end{equation}
From~(\ref{eq1h}) and using~(\ref{pot1h1})  it follows that identifying  the coefficients of $e^{-n\varphi}$ for $n=0,1,\dots, N$ implies the recursion relation
\begin{equation}\label{eq1hv}\everymath{\displaystyle}
h'_{n+2}-(n+2)h_{n+2}\,=\,-\frac{b}{2} \Big[v_n
+\sum_{j+k=n+2,\,j,k\geq2}\Big((h'_j-(j-1)h_j)(h'_k-(k-1)h_k)-h_jh_k\Big)\Big].
\end{equation}
Furthermore, identifying  the coefficients of $e^{-n\varphi}$ for $n> N$ leads to  the recursion relation
\begin{equation}\label{eq1hh}\everymath{\displaystyle}
h'_{n+2}-(n+2)h_{n+2}\,=
-\frac{b}{2}\sum_{j+k=n+2,\,j,k\geq2}\big((h'_j-(j-1)h_j)(h'_k-(k-1)h_k)-h_jh_k\big).
\end{equation}

The equations~(\ref{eq1hv}) and~(\ref{eq1hh}) are nonhomogeneous
linear ordinary differential equations with constant coefficients
for $h_{n+2}$. The nonhomogeneous terms depend on the
coefficients $h_j$ with $j=2,\dots,n$ and  on the polynomial coefficients
$v_n(\varphi)$ with $n=0,\dots,N$ of the potential function (\ref{pot10}). Therefore, it follows that  the
coefficients $h_n$ of the series~(\ref{hpot100}) can be
recursively determined as polynomials in $\varphi$. The recursion relations can be formally solved
and provide us with the polynomial solutions
\begin{equation}\label{eq1hhsol1}\everymath{\displaystyle}\begin{array}{l}
h_{n+2}(\varphi)\,=\frac{b}{2}e^{(n+2)\varphi}\int_{\varphi}^{\infty}e^{-(n+2)s}v_n(s)\\  \\
+\frac{b}{2}e^{(n+2)\varphi}\int_{\varphi}^{\infty}e^{-(n+2)s}\sum_{j+k=n+2,\,j,k\geq2}
\big((h'_j(s)-(j-1)h_j(s))(h'_k(s)-(k-1)h_k(s))-h_j(s)h_k(s)\big)ds,
\end{array}\end{equation}
for $n=0,1,\dots,N$ and
\begin{equation}\label{eq1hhsol2}\everymath{\displaystyle}\begin{array}{l}
h_{n+2}(\varphi)\,=\\  \\
\frac{b}{2}e^{(n+2)\varphi}\sum_{j+k=n+2,\,j,k\geq2}\int_{\varphi}^{\infty}e^{-(n+2)s}
\big((h'_j(s)-(j-1)h_j(s))(h'_k(s)-(k-1)h_k(s))-h_j(s)h_k(s)\big)ds,
\end{array}\end{equation}
for $n>N$.

In this way we have proved that  the differential
equation~(\ref{e1}) for potentials  $v(\varphi)$ of the
form~(\ref{pot10}) admits a (formal) one-parameter family of
psi series solutions of the form
\begin{equation}\label{hpot1002}
h(\varphi)\,=\,\frac{e^{\varphi}}{b}\,+\,\sum_{n=2}^{\infty}h_n(\varphi)e^{-(n-1)\varphi}.
\end{equation}

The first few equations  (\ref{eq1hv}) for $N\geq 2$ are
\begin{equation}\everymath{\displaystyle}
\begin{array}{rcl}
h_2'-2h_2&=&-\frac{b}{2}v_0,\\  \\
h_3'-3h_3&=&-\frac{b}{2}v_1,\\  \\
h_4'-4h_4&=&-\frac{b}{2}\left[v_2+(h_2')^2-2h_2h_2'\right].
\end{array}
\end{equation}
The corresponding solutions~(\ref{eq1hhsol1}) are
\begin{equation}\everymath{\displaystyle}
\begin{array}{rcl}
h_2(\varphi)&=&\frac{b}{2}e^{2\varphi}\int_{\varphi}^{\infty}v_0(s)e^{-2s}ds,\\  \\
h_3(\varphi)&=&\frac{b}{2}e^{3\varphi}\int_{\varphi}^{\infty}v_1(s)e^{-3s}ds,\\  \\
h_4(\varphi)&=&\frac{b}{2}e^{4\varphi}\int_{\varphi}^{\infty}v_2(s)e^{-4s}ds\,+\,
       \frac{b^3}{8}e^{4\varphi}\int_{\varphi}^{\infty}v_0(s)^2e^{-4s}ds\\  \\
       & &-\,\frac{b^3}{4}e^{4\varphi}\int_{\varphi}^{\infty}e^{-2s_1}v_0(s_1)\left(\int_{s_1}^{\infty}
          e^{-2s_2}v_0(s_2)ds_2\right)ds_1.
\end{array}
\end{equation}

\subsection{Polynomial  potentials}
In the polynomial case of (\ref{pot10})
$v(\varphi)=v_0(\varphi)$  with  $v_0(\varphi)$  being a
polynomial of degree $d$, the  family of psi series (\ref{hpot1002}) reduces
to the form
\begin{equation}\label{hpol}
h(\varphi)\,=\,\frac{e^{\varphi}}{b}+\sum_{n=1}^{\infty}b^{2n-1}\gamma_n(\varphi)e^{-(2n-1)\varphi},
\end{equation}
where $\gamma_1$ is the unique  polynomial of degree $d$ which satisfies the equation
\begin{equation}\label{eq1hv0}
\gamma'_1-2\gamma_1\,=\,-\frac{1}{2}v_0,
\end{equation}
 and the coefficients $\gamma_n(\varphi)$ $(n\geq2)$ are polynomials of degree $nd-1$, independent of the parameter $b$, which can be recursively determined by
 \begin{equation}\label{eq1gammaeven}\everymath{\displaystyle}\begin{array}{l}
\gamma'_{n+1}-2(n+1)\gamma_{n+1} \\ \\
\qquad
=\,-\frac{1}{2}\sum_{j+k=n+1,\,j,k\geq1}\big((\gamma'_j-(2j-1)\gamma_j)(\gamma'_k-(2k-1)\gamma_k)-\gamma_j\gamma_k\big).
\end{array}\end{equation}

Indeed,  if we set $n=2m-1$ ($m\geq1$) in~(\ref{eq1hh}) we have that
$$\everymath{\displaystyle}\begin{array}{l}
h'_{2m+1}-(2m+1)h_{2m+1}\\  \\
\qquad=\,-\frac{b}{2}\sum_{j+k=2m+1,\,j,k\geq2}\big((h'_j-(j-1)h_j)(h'_k-(k-1)h_k)-h_jh_k\big).
\end{array}$$
Now, taking into account~(\ref{pot1h1}) and applying induction in $m$ it is clear that
\begin{equation}\label{hodd1}
h_{2m+1}\equiv0\quad \mbox{for all}\quad m\geq0.
\end{equation}
In order to make explicit the dependence of $h(\varphi)$ on the
arbitrary parameter $b$, we introduce the functions
$\gamma_n(\varphi):=h_{2n}(\varphi)/b^{2n-1}$ . Thus, for $n=0$
equation~(\ref{eq1hv})  reduces to~(\ref{eq1hv0})
and~(\ref{eq1hh}) becomes the recursion relation
(\ref{eq1gammaeven}). We notice that the coefficients
$\gamma_n(\varphi)$ are independent of $b$.
The first few
equations ~(\ref{eq1gammaeven}) are
\begin{eqnarray}
\gamma'_2-4\gamma_2&=&-\frac{1}{2}\left[(\gamma'_1)^2-2\gamma_1\gamma'_1\right] \label{polh2},\\ \nonumber \\
\gamma'_3-6\gamma_3&=&-\left[\gamma'_1\gamma'_2-\gamma_1\gamma'_2-3\gamma'_1\gamma_2+2\gamma_1\gamma_2\right], \label{polh3}\\ \nonumber \\
\gamma'_4-8\gamma_4&=&-\frac{1}{2}\Big[2\gamma'_1\gamma'_3+(\gamma'_2)^2-2\gamma_1\gamma'_3-6\gamma_2\gamma'_2-10\gamma'_1\gamma_3 \label{polh4}\\ \nonumber
                     & &\qquad +8\gamma_1\gamma_3+8\gamma_2^2\Big].
\end{eqnarray}
In particular, (\ref{eq1hv0}) and (\ref{polh2}) imply the following explicit relations for general polynomial potentials $v_0(\varphi)$

\begin{eqnarray}\everymath{\displaystyle}
 \gamma_1(\varphi)&=&\frac{1}{2}e^{2\varphi}\int_{\varphi}^{\infty}e^{-2s}v_0(s)ds,\label{i1}
  \\
 \gamma_2(\varphi)&=&-\frac{1}{4}e^{4\varphi}\int_{\varphi}^{\infty}e^{-2s_1}v_0(s_1)\left(\int_{s_1}^{\infty}e^{-2s_2}v_0(s_2)ds_2\right)ds_1\nonumber \\ \label{i2} \\
  &+&\frac{1}{8}e^{4\varphi}\int_{\varphi}^{\infty}e^{-4s_1}v_0(s_1)^2ds_1.\nonumber
\end{eqnarray}
From~(\ref{eq1hv0}) we have  that $\gamma_1$ is a polynomial of the same degree
$d$ as $v_0(\varphi)$. Consequently the
right-hand side of~(\ref{polh2}) is a polynomial of degree $2d-1$
and so is $\gamma_2$. Now, since the coefficient of
$\gamma_n\gamma_1$ in the right-hand side of~(\ref{eq1gammaeven})
is  given by $-2(n-1)$, which is non-zero  for $n\geq2$, then
using induction in $n$ it follows that  for $n\geq2$ the
coefficient $\gamma_n(\varphi)$ is a polynomial in $\varphi$ of
degree $nd-1$.

Next we discuss two illustrative  examples.
\subsubsection{The quadratic potential}
For the quadratic potential
\begin{equation}\label{quad}
v(\varphi)\,=\,\m^2\varphi^2,
\end{equation}
we have an expansion for $h$ of the form~(\ref{hpol})
with $\gamma_1$ satisfying~(\ref{eq1hv0})
$$\gamma'_1-2\gamma_1\,=\,-\frac{1}{2}\m^2\varphi^2.$$
Hence we obtain
\begin{equation}\label{h1quad}
\gamma_1(\varphi)\,=\,\frac{\m^2}{8}(1+2\varphi+2\varphi^2),
\end{equation}
so that the  first two terms of the psi series of $h$ are
\begin{equation}\label{hpol1}
h(\varphi)\,=\,\frac{e^{\varphi}}{b}+\frac{b\m^2}{8}(1+2\varphi+2\varphi^2)e^{-\varphi}+ \cdots.
\end{equation}

The coefficients $\gamma_n$ ($n\geq2$) in ~(\ref{hpol}) are determined by~(\ref{eq1gammaeven}). For example we obtain
$$\everymath{\displaystyle}\begin{array}{rcl}
\gamma_2(\varphi)&=&-\frac{\m^4}{1024}\Big(5+20\varphi+40\varphi^2+32\varphi^3\Big),\\  \\
\gamma_3(\varphi)&=&\frac{\m^6}{1990656}\Big(703+4218\varphi+12654\varphi^2+18504\varphi^3+7344\varphi^4-5184\varphi^5\Big).
\end{array}$$

\subsubsection{The Higgs potential}

The KD period for the inflaton model with a Higgs potential
\begin{equation}\label{Higgs}
v(\varphi)=g^2(\varphi^2-\lambda^2)^2,
\end{equation}
has been studied in~\cite{DE10}, where approximate expressions for the inflaton field and the Hubble parameter  as functions of $t$ have been obtained.
According to our general result for models with polynomial potentials, the Higgs model admits a psi series of the form~(\ref{hpol}). In particular, one easily finds that
the first coefficients of this psi-series
are given by
\begin{eqnarray}\label{Higgsh1}
\gamma_1(\varphi)&=&\frac{g^2}{8}\left[3+6\varphi+6\varphi^2+4\varphi^3+2\varphi^4
-2\lambda^2(1+2\varphi+2\varphi^2)+2\lambda^4\right],\\ \nonumber \\
\nonumber\gamma_2(\varphi)&=&\displaystyle\frac{g^4}{16384}
 \Big[\lambda ^6 (1024 \varphi +768)+\lambda ^4 \left(-3072
   \varphi ^3-4864 \varphi ^2-3968 \varphi -1760\right)\\ \nonumber \\
\label{Higgsh2}   & &\;+\lambda ^2\left(3072 \varphi ^5+7424 \varphi ^4+10496 \varphi ^3+9408 \varphi
   ^2+4704 \varphi +1176\right)\\ \nonumber \\
\nonumber   & &-\left(1024 \varphi ^7+3328 \varphi ^6+6528
   \varphi ^5+8928 \varphi ^4+8928 \varphi ^3+6696 \varphi ^2+3348 \varphi
   +837\right)\Big].
 \end{eqnarray}

\subsection{ Starobinsky's  potentials}

We now consider the class of potentials of the form  (\ref{pot20}).
It includes as a particular case the Starobinsky model~\cite{STA80,STA83,WH84}
\begin{equation}\label{sta0}
v(\varphi)\,=\,\lambda(1-e^{-\alpha\varphi})^2.
\end{equation}
It can be proved  \cite{SEP} that
 the existence of  approximate solutions~(\ref{h1}) of Eq~(\ref{e1})  is only possible if
\begin{equation}\label{Napot2}
N\,\alpha\,>-\,2,
\end{equation}
so that we  will henceforth assume that (\ref{Napot2}) is satisfied.

We look for  psi series solutions of~(\ref{e1}) of the form
\begin{equation}\label{pot2exph}
h(\varphi)\,=\,\frac{e^{\varphi}}{b}+\sum_{n=1}^{\infty}h_n(u)e^{-(n-1)\varphi},
\end{equation}
where the coefficients $h_n$ are polynomials in
\begin{equation}\label{pot2x}
u:= e^{-\alpha\varphi}.
\end{equation}

If we substitute the series ~(\ref{pot2exph}) into~(\ref{e1}) we
get
\begin{equation}\label{e1pot2}\everymath{\displaystyle}
\begin{array}{l}
\frac{2}{b}\sum_{n=1}^{\infty}(\alpha u h_n'+nh_n)e^{-(n-2)\varphi}\,=\,w(u)\\  \\
+\sum_{n=2}^{\infty}\left(\sum_{j+k=n}(\alpha u h'_j+(j-1)h_j)(\alpha uh'_k+(k-1)h_k)-h_jh_k\right)e^{-(n-2)\varphi},
\end{array}\end{equation}
where
\begin{equation}\label{pot2w}
w(u):=\,\sum_{n=0}^Nv_nu^n.
\end{equation}

Since we assume that $\alpha$ is an irrational number, the powers of $u=e^{-\alpha\varphi}$ and the powers  of $e^{\varphi}$ are linearly independent functions, consequently the coefficients of $e^{-k\varphi}$ for $k\geq-1$ in both sides of~(\ref{e1pot2}) must be equal.  Then, for $k=-1$ we obtain
$ \alpha u h'_1+h_1=0$
and, since $h_1$ is a polynomial in $u$, we have that
\begin{equation}\label{pot2h1}
h_1\equiv0.
\end{equation}
Using~(\ref{pot2h1}) for $k=0$ we get
\begin{equation}\label{pot2diffh2}
 \alpha u h'_2+2h_2=\frac{b}{2}w(u).
\end{equation}

Equation~(\ref{pot2diffh2}) has a polynomial solution of degree
$N$. Proceeding in the same way, vanishing the coefficients of
$e^{-n\varphi}$ in~(\ref{e1pot2}) with $n\geq 1$, we obtain the
recurrence relation
\begin{equation}\label{pot2rech}\everymath{\displaystyle}\begin{array}{l}
\alpha u h'_{n+2}+(n+2)h_{n+2}\\  \\
\quad =\frac{b}{2}\sum_{j+k=n+2,\,j,k\geq2}\left((\alpha u h'_j+(j-1)h_j)(\alpha u h'_k+(k-1)h_k)-h_jh_k\right).
\end{array}\end{equation}
From~(\ref{pot2h1}), (\ref{pot2diffh2}) and~(\ref{pot2rech}) we conclude
that all the coefficients $h_n(u)$ in~(\ref{pot2exph}) are recursively determined by~(\ref{pot2diffh2}) and~(\ref{pot2rech}) as polynomials in $u$ .
Moreover, if we set $n=2m-1$ in~(\ref{pot2rech}), then  from~(\ref{pot2h1}) and applying induction in $m$, it follows  immediately  that
$$ h_{2m+1}\equiv0,\quad m=0,1,2,\dots.$$

To make explicit the dependence of $h(\varphi)$ on the arbitrary parameter $b$, we introduce the functions
\begin{equation}\label{pot2gammas}
\gamma_n(u):=\frac{h_{2n}(u)}{b^{2n-1}}.
\end{equation}
Thus, equations~(\ref{pot2diffh2}) and~(\ref{pot2rech}) reduce to
\begin{equation}\label{pot2diffgamma1}
  \alpha u \gamma'_1+2\gamma_1=\frac{1}{2}w(u),
\end{equation}
and
\begin{equation}\label{pot2recgamma}\everymath{\displaystyle}\begin{array}{l}
\alpha u \gamma'_{n+1}+2(n+1)\gamma_{n+1}\\ \\=
\frac{1}{2}\sum_{j+k=n+1,\,j,k\geq1}\left(( \alpha
u\gamma'_j+(2j-1)\gamma_j)( \alpha u
\gamma'_k+(2k-1)\gamma_k)-\gamma_j\gamma_k\right),\end{array}
\end{equation}
respectively. Here primes indicate  derivatives with respect to $u$.

Applying induction with respect to $n$ in~(\ref{pot2recgamma}) we easily conclude that $\gamma_n(u)$ is a polynomial in $u$ of degree at most $nN$  (in particular $\gamma_1$ is a polynomial of degree $N$).
Therefore we have proved the existence of a psi series expansion of the form
\begin{equation}\label{pot20hub}
h(\varphi)\,=\,\frac{e^{\varphi}}{b}\left(1+\sum_{n=1}^{\infty}b^{2n}\gamma_n(e^{-\alpha\varphi})e^{-2n\varphi}\right).
\end{equation}

For example,  we get the following explicit relations for  $\alpha>0$

\begin{eqnarray}\everymath{\displaystyle}
 \gamma_1(u)&=&\frac{1}{2\alpha}u ^{-2/\alpha}\int_0^u s^{2/\alpha-1}w(s)ds,\label{is1}
  \\
 \gamma_2(u)&=&-\frac{1}{4\alpha^2}u^{-4/\alpha}\int_0^us_1^{2/\alpha-1}w(s_1)
 \left(\int_0^{s1}s_2^{2/\alpha-1}w(s_2)ds_2\right)ds_1\nonumber \\ \label{is2} \\
  &+&\,\frac{1}{8\alpha}u^{-4/\alpha}\int_0^us_1^{4/\alpha-1}w(s_1)^2ds_1,
\nonumber
\end{eqnarray}
For $\alpha<0$ the same expressions hold with the lower
limits of the integrals substituted by $\infty$.

\subsubsection{The Starobinsky model}

The potential function of the  Starobinsky model  is given by
\begin{equation}\label{sta}
v(\varphi)\,=\,\lambda\,(1-e^{-\alpha\varphi})^2,\quad \alpha>-1.
\end{equation}
Then the corresponding equation~(\ref{pot2diffgamma1}) reduces to
\begin{equation}\label{ss1}
 \alpha u \gamma'_1+2\gamma_1\,=\,\frac{\lambda}{2}(1-u)^2,
\end{equation}
and the first two equations~(\ref{pot2recgamma}) are
$$\everymath{\displaystyle}\begin{array}{rcl}
 \alpha u \gamma'_2+4\gamma_2&=&\frac{( \alpha u )^2}{2}(\gamma'_1)^2+ \alpha u \gamma_1\gamma'_1,\\  \\
 \alpha u \gamma'_3+6\gamma_3&=&( \alpha u )^2\gamma'_1\gamma'_2+3 \alpha u \gamma'_1\gamma_2+ \alpha u \gamma_1\gamma'_2+2\gamma_1\gamma_2.
\end{array}$$
Therefore, the first polynomial coefficients of the expansion~(\ref{pot2exph}) turn out to be given by
\begin{equation}\label{stagammas}\everymath{\displaystyle}\begin{array}{rcl}
\frac{\gamma_1(u)}{\lambda}&=&\frac{u^2}{4 ( \alpha +1)}-\frac{u}{ \alpha +2}+\frac{1}{4},\\  \\
\frac{\gamma_2(u)}{\lambda^2}&=&\frac{ \alpha u ^4}{32 (\alpha+1)^2}-\frac{\alpha (2 \alpha+3) u^3}{4 (\alpha+1) (\alpha+2) (3 \alpha+4)}\\  \\
           & &+\frac{\alpha (5  \alpha +6) u^2}{16 ( \alpha +1) ( \alpha +2)^2}-\frac{ \alpha u }{4 ( \alpha +2)( \alpha +4)},\\  \\
\frac{\gamma_3(u)}{\lambda^3}&=&\frac{\alpha (4\alpha+1) u^6}{384 (\alpha+1)^3}-\frac{\alpha (4 \alpha+5)
               \left(9 \alpha^2+16 \alpha+4\right) u^5}{32 (\alpha+1)^2 (\alpha+2) (3\alpha+4) (5 \alpha+6)}\\  \\
           & &+\frac{ \alpha  \left(162  \alpha ^4+739  \alpha^3+1190  \alpha^2+776  \alpha+160\right)u^4}{128 ( \alpha +1)^2 ( \alpha +2)^2 (2  \alpha +3) (3  \alpha +4)}\\  \\
           & &-\frac{ \alpha  \left(42  \alpha^4+301 \alpha^3+674  \alpha^2+584 \alpha+160\right) u^3}{48 ( \alpha+1) ( \alpha+2)^3 ( \alpha+4) (3 \alpha +4)}\\  \\
           & &+\frac{ \alpha \left(13  \alpha^2+58  \alpha +40\right) u^2}{64 ( \alpha+2)^2( \alpha +3) ( \alpha+4)}-\frac{ \alpha u }{16 ( \alpha+4) ( \alpha +6)}.
\end{array}\end{equation}

\subsection{Logolinear series}

Once we have determined the psi series~(\ref{hpot1002}),~(\ref{hpol}) and~(\ref{pot20hub}) for $h(\varphi)$, then from~(\ref{e}) psi series of logolinear type depending on the variable $t$
~\cite{LA05,DE10,HAN14,HAN19}  for the inflaton field $\varphi$ and the reduced Hubble parameter $h$ can be derived.
Thus, for polynomial potentials, if we insert ~(\ref{hpol}) into~(\ref{e}) we get
\begin{equation}\label{polvarphi}
\dot{\varphi}\,=\,-\frac{e^{\varphi}}{b}-\sum_{n=1}^{\infty}b^{2n-1}
\left(\gamma_n'(\varphi)-(2n-1)\gamma_n(\varphi)\right)e^{-(2n-1)\varphi}.
\end{equation}
It can be easily checked that~(\ref{polvarphi}) admits a family of formal psi series solutions of the form
\begin{equation}\label{psivarphi1}
\varphi(t)\,=\,-\,x\,+\,\sum_{n=1}^{\infty}\alpha_n(x)(t-t^*)^{2n},
\end{equation}
where
\begin{equation}\label{equis}
x:=\log\left(\frac{t-t^*}{b}\right),
\end{equation}
and $\alpha_n(x)$ can be recursively determined as polynomials in $x$. Next we sketch  the main ideas of the proof. First, we introduce the polynomials $\theta_n(\varphi)$ and their coefficients $\theta_{n,j}$, $j=0,\dots,nd$
through
$$\theta_n(\varphi)\,:=\,\gamma_n'(\varphi)-(2n-1)\gamma_n(\varphi)\,=\,\sum_{j=0}^{nd}\theta_{n,j}\varphi^j,$$
and then we rewrite~(\ref{polvarphi}) as
\begin{equation}\label{ppsi1}
\dot{\varphi}\,=\,-\frac{e^{\varphi}}{b}-\sum_{n=1}^{\infty}\sum_{j=0}^{nd}\theta_{n,j}\varphi^jb^{2n-1}e^{-(2n-1)\varphi}.
\end{equation}
and replace~(\ref{psivarphi1}) into~(\ref{ppsi1}). For the left hand side we have
\begin{equation}\label{ppsi2}
\dot{\varphi}\,=\,-\frac{1}{t-t^*}+\sum_{n=1}^{\infty}[\alpha_n'(x)+2n\alpha_n(x)](t-t^*)^{2n-1}.
\end{equation}
In order to expand the right hand side of~(\ref{ppsi1}) in odd powers of $(t-t^*)$ we introduce the Bell's polynomials~\cite{B34} defined through
\begin{equation}\label{Bell}
\exp\left(\sum_{n=1}^{\infty}x_nz^n\right)\,=\,\sum_{n=0}^{\infty}C_n(x_1,\dots,x_n)z^n.
\end{equation}
Then we have that
\begin{equation}\label{ppsi3}
b^{2n-1}e^{-(2n-1)\varphi}\,=\,(t-t^*)^{2n-1}\sum_{m=0}^{\infty}C_{m}(A_{m,n}(x))(t-t^*)^{2m},
\end{equation}
where we are introducing the vectorial functions
$$A_{m,n}(x)\,:=\,-(2n-1)\Big(\alpha_1(x),\dots,\alpha_m(x)\Big).$$
Thus, by replacing~(\ref{ppsi2}) and~(\ref{ppsi3}) into~(\ref{ppsi1}) and equating the coefficients of $(t-t^*)^{2n-1}$ we obtain the equations
\begin{equation}\label{pot1eqaseven}\everymath{\displaystyle}\begin{array}{lll}
\alpha'_n+(2n+1)\alpha_n&=&\alpha_n-C_n({A}_{n,0})-\\  \\
 & &\mbox{coeff}\left[\sum_{m=1}^{n}\sum_{j=0}^{md}\sum_{l=0}^{n-m}\theta_{m,j}C_l({A}_{l,m})
\left(-x+\sum_{k=1}^{n-1}\alpha_k\tau^{2k}\right)^j,\tau^{2(n-m-l)}\right].
\end{array}\end{equation}
The first equation~(\ref{pot1eqaseven}) takes the form
$$\alpha_1'+3\alpha_1\,=\,-\theta_1(-x),$$
so that $\alpha_1$ can be determined as a polynomial of degree $d$. Then, since~(\ref{pot1eqaseven}) is an ordinary linear differential equation for $\alpha_n$ with nonhomogeneus term depending only on $\alpha_1,\dots,\alpha_{n-1}$, then using induction in $n$ it follows that the coefficients $\alpha_n(x)$ can be recursively determined as polynomials of degree at most $nd$. For instance, equation~(\ref{pot1eqaseven}) for $n=2$ is
$$
\alpha_2'+5\alpha_2\,=\,-\frac{\alpha_1(x)^2}{2}+\alpha_1(x)[\theta_1(-x)-\theta_1'(-x)]-\theta_2(-x).
$$
Furthermore, by substituting ~(\ref{psivarphi1}) into~(\ref{hpol}), we find an expansion
\begin{equation}\label{psihub1}
h(t)\,=\,\frac{1}{t-t^*}\,+\,\sum_{n=1}^{\infty}\beta_n(x)(t-t^*)^{2n-1},
\end{equation}
where $\beta_n(x)$ are polynomials in $x$ that can be written in terms of $\alpha_k$, $\gamma_k$, $k=1,\dots,n$.
For example, we have that
$$\everymath{\displaystyle}\begin{array}{lll}
\beta_1(x)&=&\alpha_1(x)+\gamma_1(-x),\\  \\
\beta_2(x)&=&\alpha_2(x)+\frac{\alpha_1(x)^2}{2}+\alpha_1(x)[\gamma_1'(-x)-\gamma_1(-x)]+\gamma_2(-x).
\end{array}$$
The psi series~(\ref{psivarphi1})-(\ref{psihub1}) are the logolinear expansions derived in~\cite{HAN19} for inflaton models with polynomial potentials in a  flat universe.

Analogously, for the Starobinsky potentials~(\ref{pot20}), replacement of~(\ref{pot20hub}) into~(\ref{e})
leads us to
\begin{equation}\label{Stavarphi}
\dot{\varphi}\,=\,-\frac{e^{\varphi}}{b}\,+\,\sum_{n=1}^{\infty}b^{2n-1}\Big(\alpha e^{-\alpha\varphi}
\gamma_n'(e^{-\alpha\varphi})+(2n-1)\gamma_n(e^{-\alpha\varphi})\Big)e^{-(2n-1)\varphi}.
\end{equation}
It can be proved that~(\ref{Stavarphi}) admits a family of psi series solutions of the form
\begin{equation}\label{psivarphi2}
\varphi(t)\,=\,-\log\left(\frac{t-t^*}{b}\right)\,+\,\sum_{n=1}^{\infty}\alpha_n(\sigma)(t-t^*)^{2n},
\end{equation}
where
\begin{equation}\label{sigma}
\sigma:=\left(\frac{t-t^*}{b}\right)^{\alpha},
\end{equation}
and $\alpha_n(\sigma)$ can be recursively determined as polynomials in $\sigma$. We outline the main ideas of the proof. By introducing the polynomials $\theta_n(u)$ and their coefficients $\theta_{n,j}$, $j=0,\dots,nN$ through
$$\theta_n(u)\,:=\,\alpha u\gamma_n'(u)+(2n-1)\gamma_n(u)\,=\,\sum_{j=0}^{nN}\theta_{n,j}u^j,$$
the equation~(\ref{Stavarphi}) takes the form
\begin{equation}\label{spsi1}
\dot{\varphi}\,=\,-\frac{e^{\varphi}}{b}+\sum_{n=1}^{\infty}\theta_{n,j}\,b^{2n-1}e^{-(2n-1+j\alpha)\varphi}.
\end{equation}
The left hand side of~(\ref{spsi1}) is expanded in odd powers of $(t-t^*)$ as
\begin{equation}\label{spsi2}
\dot{\varphi}\,=\,-\,\frac{1}{t-t^*}+\sum_{n=1}^{\infty}\Big(\alpha\sigma\alpha_n'(\sigma)+2n\alpha_n(\sigma)\Big)(t-t^*)^{2n-1}.
\end{equation}
By using the Bell's polynomials~(\ref{Bell}) we can also expand the terms in the series of the right hand side of~(\ref{spsi1}) as
\begin{equation}\label{spsi3}
b^{2n-1}e^{-(2n-1+j\alpha)\varphi}\,=\,(t-t^*)^{2n-1}\sigma^j\sum_{m=0}^{\infty}C_m(A_{m,n,j}(\sigma))(t-t^*)^{2m},
\end{equation}
where the vectorial polynomials $A_{m,n,j}$ have been defined as
$$A_{m,n,j}(\sigma)\,:=\,-(\alpha j+2n-1)\Big(\alpha_1(\sigma),\dots,\alpha_n(\sigma)\Big).$$
Thus, by replacing~(\ref{spsi2}) and~(\ref{spsi3}) into~(\ref{spsi1}), taking into account that as $\alpha$ is an irrational number, the powers of $\sigma$ and the powers of $(t-t^*)$ are linearly independent, then equating the coefficients of $(t-t^*)^{2n-1}$ in both sided of the equations, we obtain that $\alpha_n$ satisfies the differential equation
\begin{equation}\label{pot2rec2}\everymath{\displaystyle}
a\sigma\alpha'_{n}+(2n+1)\alpha_{n}\,=\,\alpha_{n}-C_{n}(A_{n,0,0})+
\sum_{k+m=n,\,k\geq1,m\geq0}\sum_{j=0}^{kN}\theta_{k,j}\sigma^jC_m(A_{m,k,j}(\sigma)).
\end{equation}
As the right hand side of~(\ref{pot2rec2}) depends only on $\alpha_j$, $j=1,\dots,n-1$, this equation shows that the coefficients $\alpha_n(\sigma)$ can be recursively obtained as polynomials of degree at most $nN$.
Thus, for instance we have that the two first equations~(\ref{pot2rec2}) are
$$\everymath{\displaystyle}\begin{array}{lll}
\alpha\sigma\alpha'_1+3\alpha_1&=&\theta_1(\sigma),\\  \\
\alpha\sigma\alpha'_2+5\alpha_2&=&-\frac{\alpha_1^2(\sigma)}{2}-\alpha_1(\sigma)\left[\alpha\sigma\theta_1'(\sigma)+\theta_1(\sigma)\right]+\theta_2(\sigma).
\end{array}$$
Furthermore, the replacement of~(\ref{psivarphi2}) into~(\ref{pot20hub}) provides us with the formal psi series for the reduced Hubble parameter
\begin{equation}\label{psihub2}
h(t)\,=\,\frac{1}{t-t^*}\,+\,\sum_{n=1}^{\infty}\beta_n(\sigma)(t-t^*)^{2n-1},
\end{equation}
where $\beta_n(\sigma)$ are polynomials in $\sigma$ that can be determined in terms of $\alpha_k$, $\gamma_k$, $k=1,\dots,n$. Thus, for example, we have that
$$\everymath{\displaystyle}\begin{array}{lll}
\beta_1(\sigma)&=&\alpha_1(\sigma)+\gamma_1(\sigma),\\  \\
\beta_2(\sigma)&=&\alpha_2(\sigma)+\frac{\alpha_1(\sigma)^2}{2}+\gamma_2(\sigma)-\left[\alpha\sigma\gamma_1'(\sigma)+\gamma_1(\sigma)\right]\alpha_1(\sigma).
\end{array}$$

We notice that for the Starobinsky model~(\ref{sta}), the psi series~(\ref{psivarphi2})-(\ref{psihub2})
correspond to the logolinear expansions   for the inflaton model determined in~\cite{HAN19} for a flat universe,
 $\lambda=\Lambda^4$ and $\alpha\,=\,\pm\frac{2}{3}$.

The determination of logolinear series for the general class of polynomial exponential potentials~(\ref{pot10}) is more involved and together with the logolinear series for some generalizations of the class of Starobinsky potentials~(\ref{pot20}) will be the subject of a future work .

\section{APPLICATIONS}\label{app}

Many analytical calculations for inflation models assume the slow-roll approximation of (\ref{eq1})-(\ref{eq2})
\begin{equation}\label{sr0}
 H^2\sim \frac{1}{3m_{Pl}^2}V,\quad \dot{\phi} \sim -\frac{1}{3H}\frac{dV}{d\phi},
\end{equation}
which in terms of the reduced Hubble function $h(\varphi)$ can be expressed as
\begin{equation}\label{sr}
h\sim \sqrt{v},\quad  h'\sim (\sqrt{v})'.
\end{equation}
However for calculations involving  contributions of the KD period it is more appropriate to use  truncations of psi series.

\subsection{The inflation region}\label{appA}
From (\ref{cl}) it follows that we  may estimate the value
$\varphi_{class}(b)$  of the inflaton field at the beginning of
the classical period as a function of $b$  by means of the
equation
\begin{equation}\label{app1}
h(\varphi_{class},b)=h_{p}=\frac{1}{8 \pi}.
\end{equation}
In particular, since the  approximation (\ref{h1}) is model independent, we obtain the general estimation
\begin{equation}\label{app1a}
\varphi_{class}\approx \log\Big(\frac{b}{8\pi}\Big) .
\end{equation}

 A standard procedure  \cite{BA09,MA18} to calculate the boundary values $\varphi_{in}$ and  $\varphi_{end}$ of the inflation period  proceeds as follows:
 we first determine   $\varphi_{end}$ by imposing $\epsilon_V=1$ where
 \begin{equation}\label{sr1}
\epsilon_V= \frac{m_{Pl}^2}{2}\Big(\frac{V_{\phi}}{V} \Big)^2,
 \end{equation}
is  the  potential slow-roll  parameter. It leads to the equation
\begin{equation}\label{app3b}
v'(\varphi_{end})=\frac{2}{\sqrt{3}}v(\varphi_{end}).
\end{equation}
Then $\varphi_{in}$ is calculated by using the slow-roll approximation during the whole period of inflation and by  adjusting the amount of inflation
\begin{equation}\label{Neb0}
N\,\approx\,\frac{1}{3}\int_{\varphi_{end}}^{\varphi_{in}} \frac{\sqrt{v(\varphi)}}{\big(\sqrt{v(\varphi)}\big)'}d\varphi
\end{equation}
to  $N\sim 60$. This method leads to the same inflation period for all the solutions of the inflaton model, since (\ref{app3b}) and (\ref{Neb0}) depend only on the potential $v(\varphi)$.
Nevertheless, in general, the inflation period depends on the solution used since  each solution has a different kinetically dominated fraction of its inflation region.

In order to include the effect of the KD part  of the inflation period,  we may proceed as follows.
We first calculate the function $\varphi_{in}(b)$
 by means of   the equation (\ref{inf})
\begin{equation}\label{app3a}
h(\varphi_{in},b)=\sqrt{\frac{3v(\varphi_{in})}{2}},
\end{equation}
where $h$ is approximated by a  truncation with a given number $n+1$ of terms of the psi series ((\ref{hpot1002}),~(\ref{hpol}) or~(\ref{pot20hub}))  corresponding to the model.
 Then, to calculate
$N(b)$  we look for a  value $\varphi^*(b)$ such that the KD
approximation holds for the interval
$\varphi\in(\varphi^*(b),\varphi_{in}(b))$,
while  the slow-roll approximation holds  for the remaining part of the inflation period. Thus, we estimate
 the \emph{amount of inflation} $N(b)$  for  the psi series corresponding to $b$ by
\begin{equation}\label{Neb}
N(b)\,\approx\,\frac{1}{3}\int_{\varphi_{end}}^{\varphi^*(b)}
\frac{\sqrt{v(\varphi)}}{\big(\sqrt{v(\varphi)}\big)'}d\varphi\,+\,\frac{1}{3}
\int_{\varphi^*(b)}^{\varphi_{in}(b)}\frac{h(\varphi,b)}{h_{\varphi}(\varphi,b)}d\varphi.
\end{equation}
The point now is how to get an appropriate value for $\varphi^*(b)$. To this end  we observe that  the KD approximation means that $h'\sim h$ and $h\gg\sqrt{v}$, and that it does not hold as $h'(\varphi)$
gets close to zero. Thus, from (\ref{e1}) we have that the truncated psi series cannot provide a good approximation when $h_{\varphi}(\varphi,b)^2$ becomes small.
Consequently, we   take $\varphi^{(n)}(b)$ ($\varphi^{(n)}(b)<\varphi_{in}(b)$) as the local minimum of $h_{\varphi}(\varphi,b)^2$ closest to $\varphi_{in}(b)$.  Numerical evidence shows that the KD approximation is not acceptable for
$\varphi$ very close to $\varphi^{(n)}(b)$, then $\varphi^*(b)$ must be strictly larger than $\varphi^{(n)}(b)$. In order to introduce this
correction we use
for $\varphi^*(b)$ an expression of the form
$$\varphi^*(b)=\alpha\varphi^{(n)}(b)+\beta\varphi_{in}(b),$$
with  two real positive parameters $\alpha$ and $\beta$ such that $\alpha+\beta=1$.
For example, we have taken $\alpha=0.9$ and $\beta=0.1$ in the examples of figures 3-7. Figures 4 and 7 show how the KD approximations fit accurately the numerical approximation for $\varphi\in(\varphi^*(b),\varphi_{in}(b))$.

The  approximation (\ref{Neb}) can be used to select the
appropriate value of the  parameter $b$, and consequently the
initial conditions, corresponding to a solution of~(\ref{e1}) with
a  previously fixed
 value of $N(b)$.

Figures 3 and 4 show the application of  this method to the
quadratic model $v(\varphi)=\m^2\varphi^2$. We use the truncated
 psi series  with 7 terms. It is worthy to point out that very good approximations are also obtained when fewer terms
 $4,5 $ or $6$ are used. We illustrate this fact in figure 3 (right) where  we use also truncated series with 4, 5 or 6 terms. It means that the KD approximation is rather reliable.

\begin{figure}[!h]\label{figq2}
\begin{center}
        \includegraphics[width=7cm]{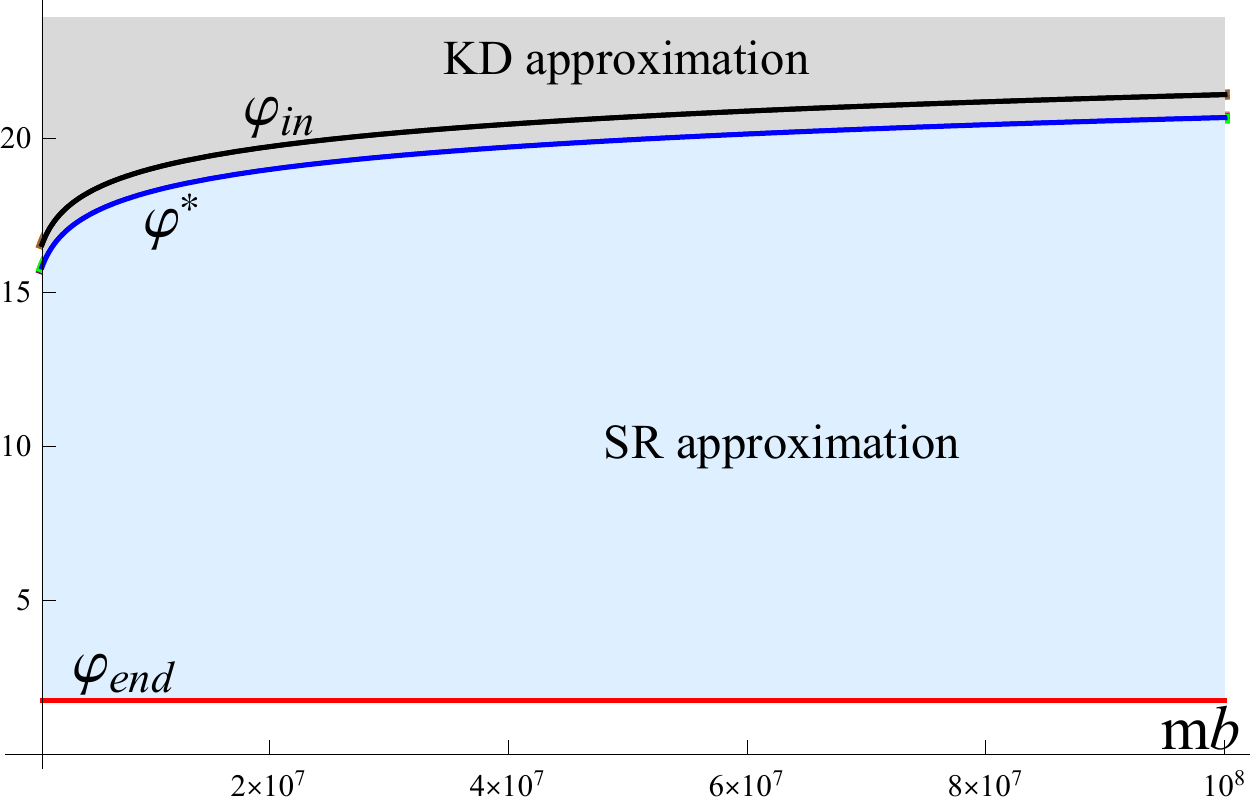}
        \hspace{1cm}
        \includegraphics[width=7cm]{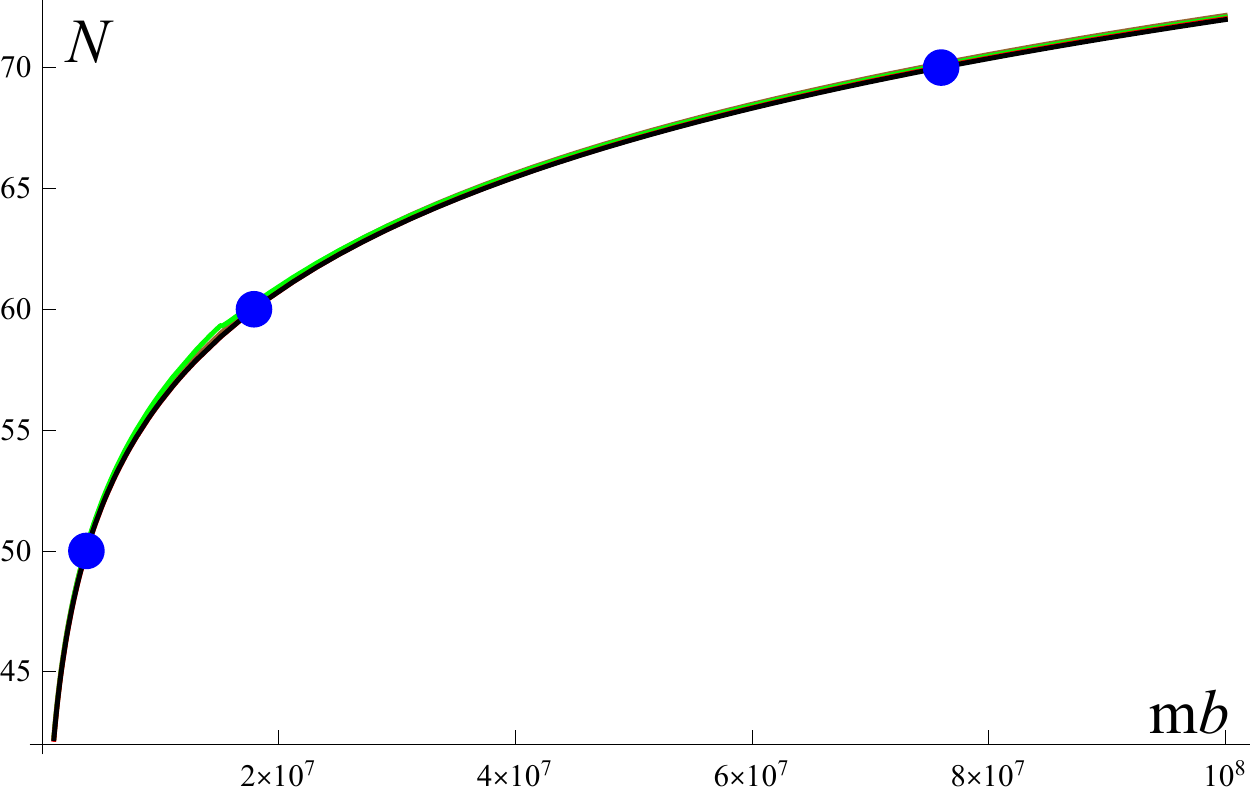}
\end{center}
    \caption{
 The left figure shows the lines $\varphi_{in}(b)$ (black line), $\varphi^*(b)$ (blue line) and $\varphi_{end}$ (red line) for the quadratic model
$v(\varphi)=\m^2\varphi^2$. It also shows  the regions in the $(\m
b, \varphi)$ plane corresponding to the KD approximation (grey
region) and the SR approximation (blue region). The right figure
shows the amount of inflation $N(b)$ corresponding to  the approximations with  4 (brown
line), 5 (green line), 6 (red line) and 7 (black line) terms, respectively,  in
the truncated series~(\ref{hpol})
provided by~(\ref{Neb}). The lines are almost completely
overlapped, which shows that a very good approximation is obtained
with only a few terms of the truncated series. The blue dots
indicate the values $N=50,\,60$ and $70$.}
\end{figure}

 In particular for $N(b)=60$ we obtain $b_{60}\,=\,\frac{\hat{b}_{60}}{\m}$, $\hat{b}_{60}\approx 1.79104\times10^{7}$.
 In figure 4, we plot the numerical solution
$h(\varphi)$ of~(\ref{e1}) with initial condition $\varphi_0=22$,
$h_0=h(\varphi_0,b_{60})$. Numerical computation of the amount of
inflation for this solution leads to $N_{num}\approx60.86$ .

\begin{figure}[!h]\label{figq3}
\begin{center}
       \includegraphics[width=8cm]{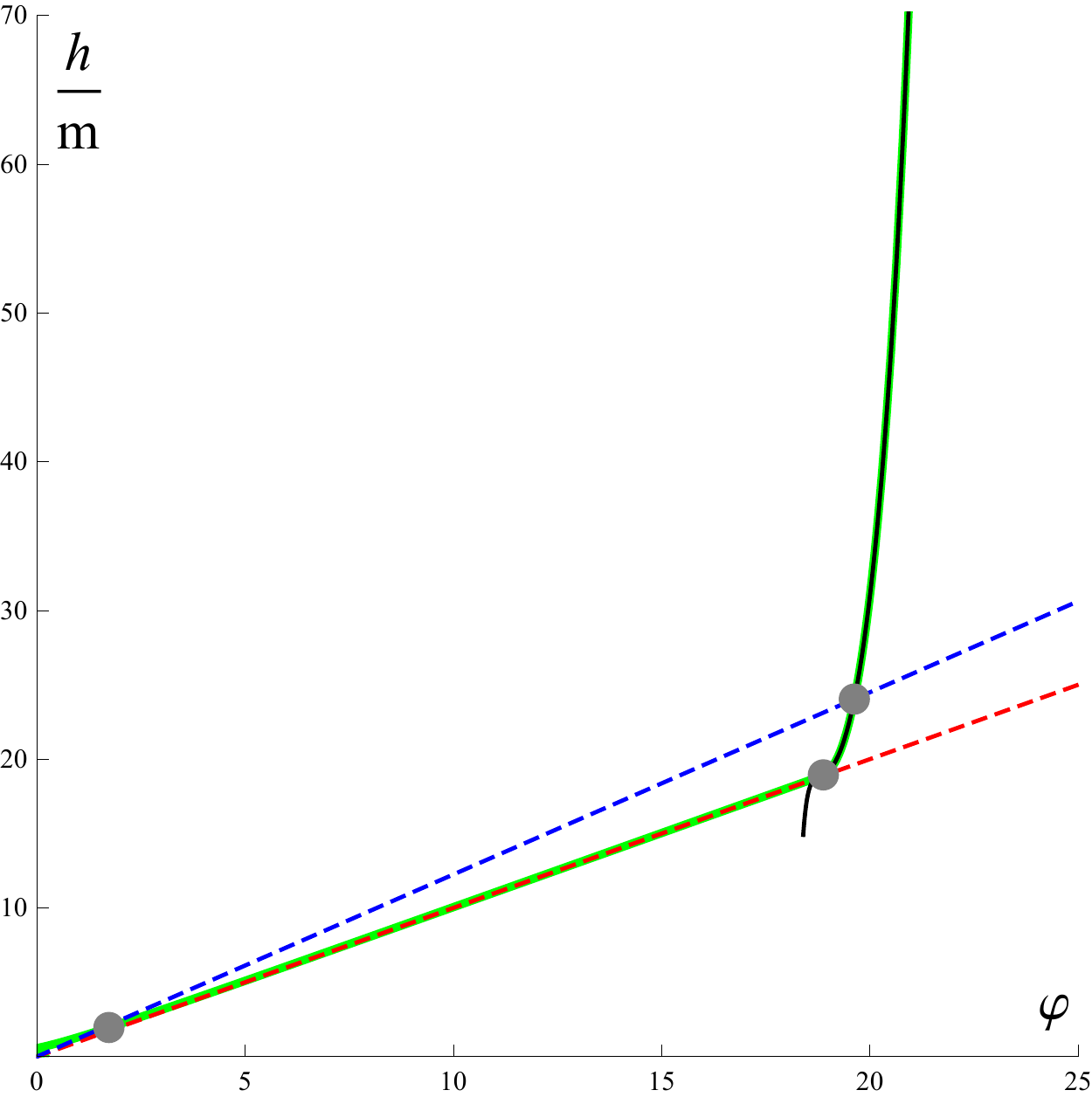}
\end{center}
\caption{The green line shows the rescaled numerical solution
$h/\m=h(\varphi)/\m$  of (\ref{e1}) with initial condition
$(\varphi_0=22, h_0=h(\varphi_0,b_{60})\approx 200.79 \m)$ for the quadratic model
$v(\varphi)=\m^2\varphi^2$.
The black line shows the KD approximation. The region between blue and red  dotted lines is
the inflation region. The grey dots correspond (from right to
left) to $\varphi_{in}$ , $\varphi^*$ and
$\varphi_{end}$. We notice that the black line (KD approximation) overlaps the green line (numerical approximation)
for $\varphi\in(\varphi^*,\varphi_{in})$, while the red dashed line (SR approximation) overlaps the green line for
$\varphi\in(\varphi_{end},\varphi^*)$.} \end{figure}

We may also consider $\varphi_{in}$ as a function of $N$,
and compare it with the SR approximation for $\varphi_{in}$ which derives from~(\ref{Neb0}) i.e.
\begin{equation}\label{insr}
\tilde{\varphi}_{in}(N)=\sqrt{3+6N}.
\end{equation}
Figure 5 shows the graphs of both approximations to $\varphi_{in}(N)$. It can be observed that the result of taking into account the effect of the KD stage leads to greater values of $\varphi_{in}(N)$.

\begin{figure}[!h]\label{figq4}
\begin{center}.
       \includegraphics[width=8cm]{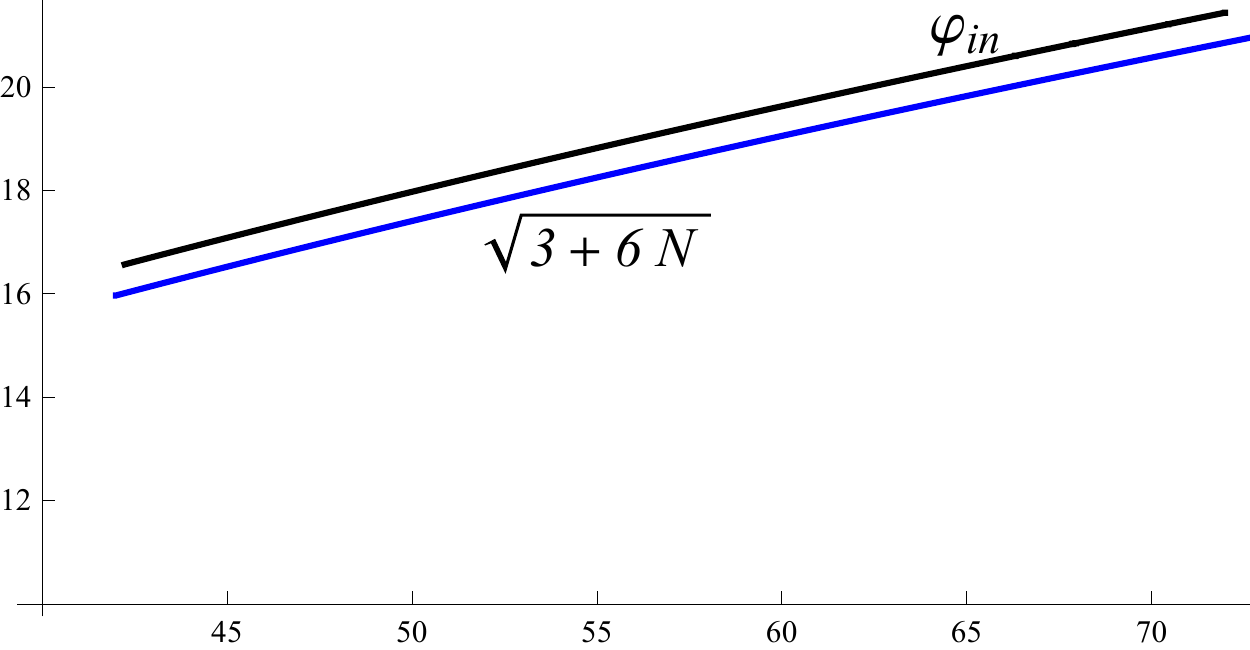}
\end{center}
\caption{The black line shows $\varphi_{in}$ as a function of $N$, where $\varphi_{in}$ is obtained from~(\ref{app3a}) and $N$ from the
formula~(\ref{Neb}), which takes into account the contributions of both the KD stage and the SR stage. The blue line shows $\varphi_{in}$ as a function
of $N$ as given by the SR approximation~(\ref{insr}).}
\end{figure}

We apply the same scheme to the Starobinsky model
$v(\varphi)=\lambda(1-e^{-\frac{\varphi}{\sqrt{3}}})^2$, and exhibit
the corresponding graphics in figures 6 and 7. We use the truncated
 psi series  with 7 terms, although we exhibit also, in figure 6 (right) how the difference with the results corresponding to truncated series with 4, 5 or 6 terms are almost imperceptible. In this case, a
solution with $N(b)=60$ corresponds to a value $b_{60}=\hat{b}_{60}/\sqrt{\lambda}$, $\hat{b}_{60}\approx 2649.03$ and the numerical
solution with initial condition $\varphi_0=12$, $h_0=h(\varphi_0,b_{60})$ gives  the number of e-folds $N_{num}(b)\approx 59.5$.

\begin{figure}[!h]\label{figS2}
\begin{center}
        \includegraphics[width=7cm]{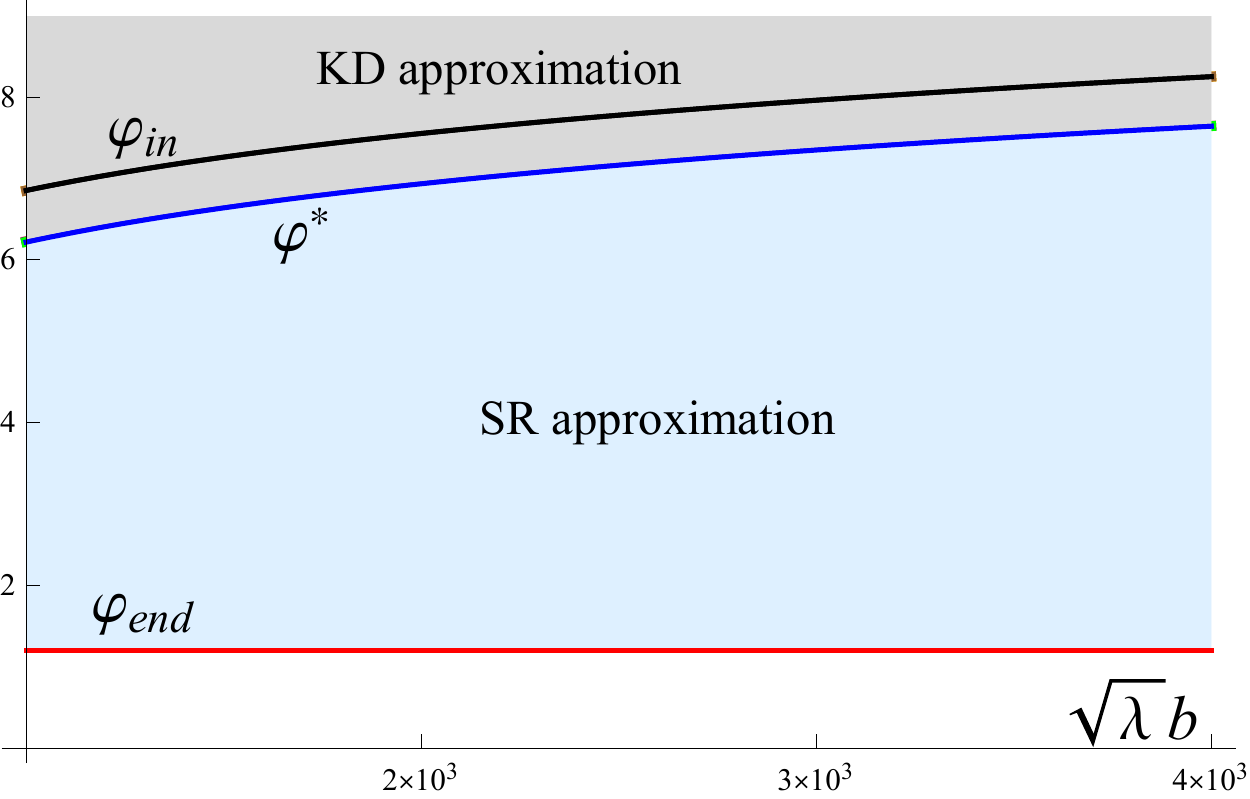}
        \hspace{1cm}
        \includegraphics[width=7cm]{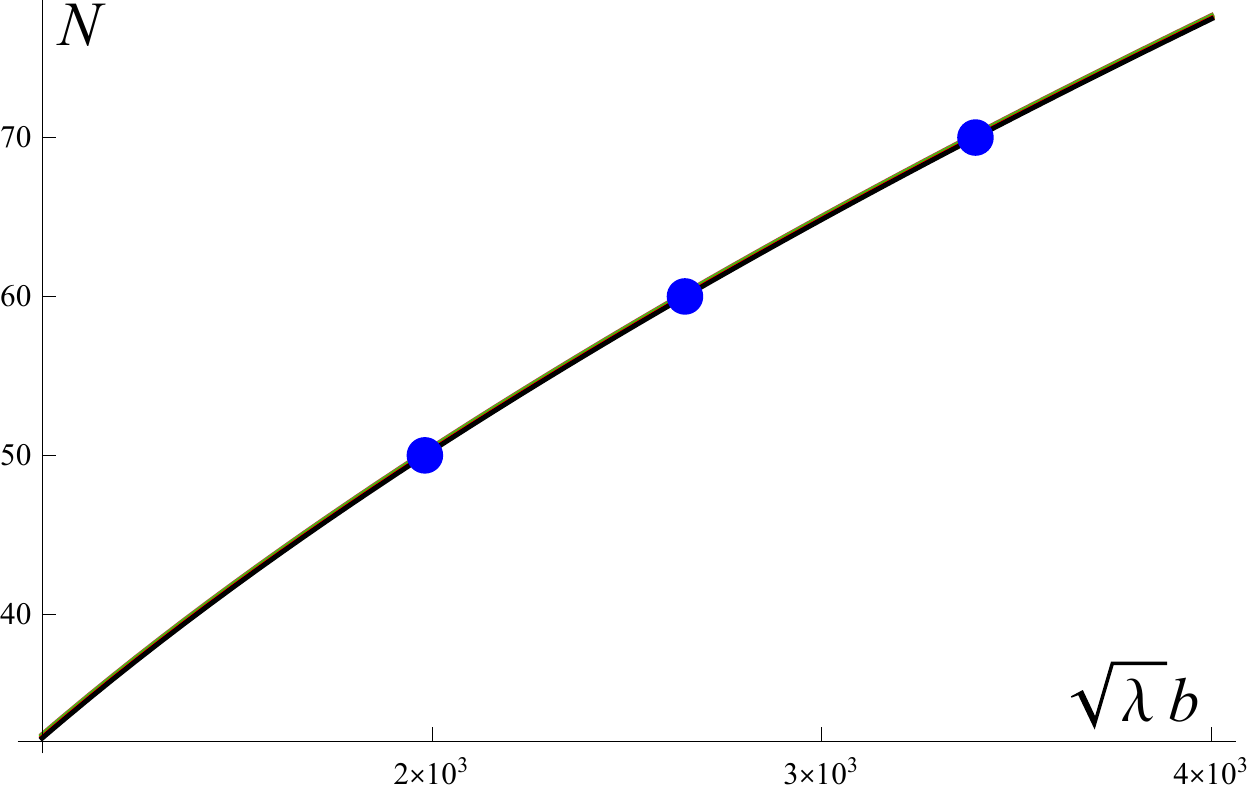}
\end{center}
    \caption{
    The left figure shows the lines $\varphi_{in}(b)$ (black line), $\varphi^*(b)$ (blue line) and $\varphi_{end}$ (red line),
    as well as the regions in the $(\sqrt{\lambda}b, \varphi)$ plane corresponding to the KD approximation (grey region) and the
    SR approximation (blue region)    for the Starobinsky model
$v(\varphi)=\lambda(1-e^{-\frac{\varphi}{\sqrt{3}}})^2$.
The right figure shows  the amount of inflation $N(b)$  determined by the approximations ~(\ref{Neb}) with 4 (brown
line), 5 (green line), 6 (red line) and 7 (black line) terms, respectively,
in the truncated series~(\ref{pot20hub})
provided by~(\ref{Neb}).
 The lines are almost completely overlapped, which shows that a very good approximation is obtained with only a few terms in the truncated series .
The blue dots correspond to $N=50,\,60$ and $70$.}
\end{figure}
\begin{figure}[!h]\label{figS3}
\begin{center}
       \includegraphics[width=7.7cm]{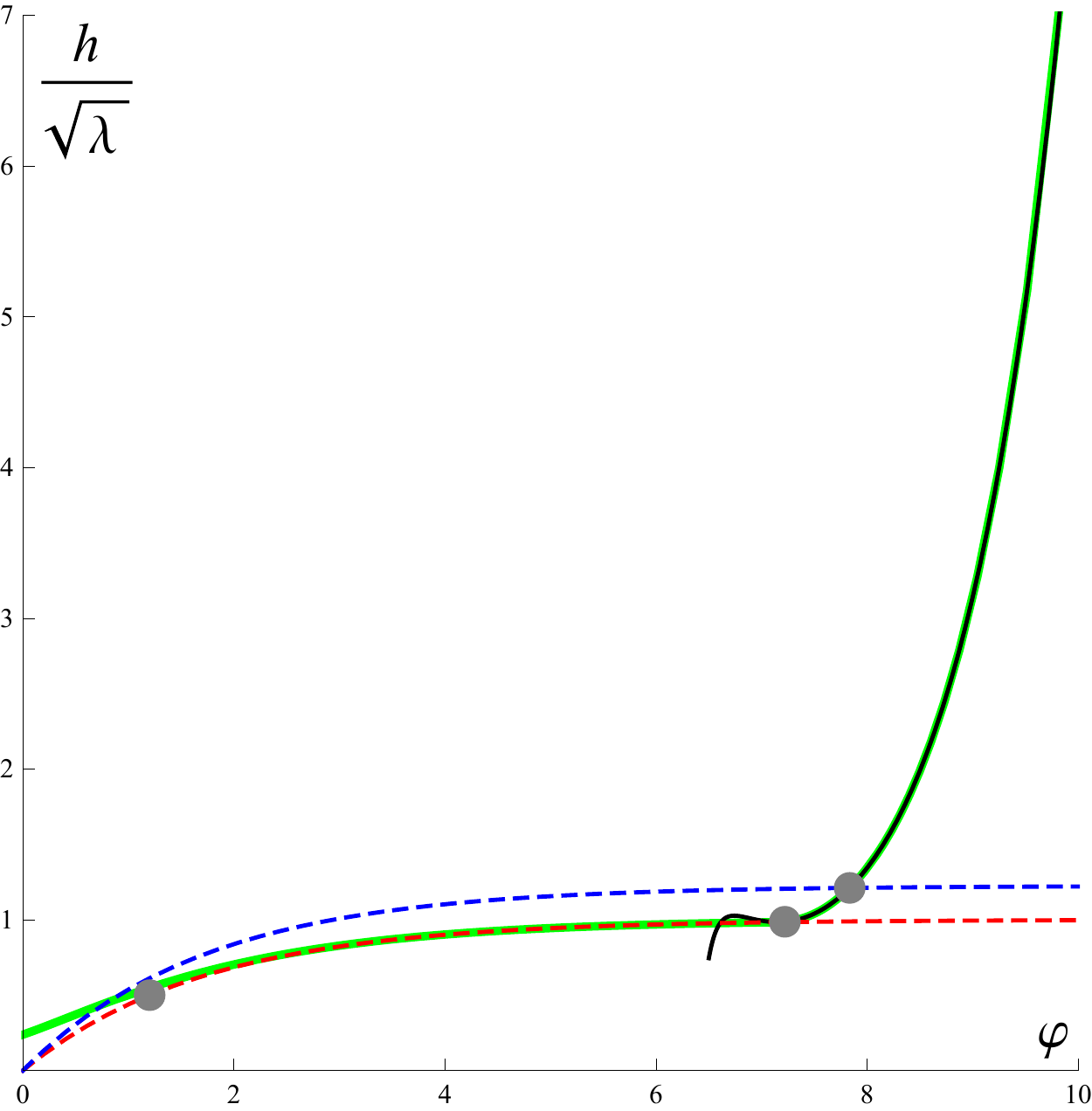}
\end{center}
\caption{The green line shows the numerical solution
$h/\sqrt{\lambda}=h(\varphi)/\sqrt{\lambda}$  of (\ref{e1}) with initial condition
$(\varphi_0=12, h_0=h(\varphi_0,b_{60})\approx 61.4435\sqrt{\lambda})$  for the Starobinsky model
$v(\varphi)=\lambda(1-e^{-\frac{\varphi}{\sqrt{3}}})^2$. The black
line shows the KD approximation. The region between blue and red
dotted lines is the inflation region. The grey dots correspond
(from right to left) to $\varphi_{in}$ ,
$\varphi^*$ and $\varphi_{end}$. It has to be noticed that the black line (KD approximation) overlaps the green line (numerical approximation) for $\varphi\in(\varphi^*,\varphi_{in})$, while the red dashed line (SR approximation) overlaps the green line for $\varphi\in(\varphi_{end},\varphi^*)$.} \end{figure}

\pagebreak

\subsection{The KD period and the  Mukhanov-Sasaki equation}
 As it was shown in ~\cite{HAN14,HAN19,DE10}  truncations of logolinear series expansions involving  powers of $t-t^*$ and $\log(t-t^*)$   are useful to determine the effect of the KD period on the power spectrum of the primordial curvature and tensor perturbations. These perturbations  are characterized by the Mukhanov-Sasaki equation \cite{LI00}
\begin{equation}\label{perct}
\Big(\frac{d^2}{d \eta^2}+k^2-W_{\alpha}(\eta)\Big) S_{\alpha}(k,\eta)=0; \quad \alpha=R,T,
\end{equation}
where the potential $W_{\alpha}(\eta)$ felt by the perturbations is
\begin{equation}\label{w2}\everymath{\displaystyle}
W_{\alpha}(\eta)=\left\{\begin{array}{c}\mbox{$\frac{Z_{\eta \eta}}{Z}$ for curvature perturbations}\\\\\mbox{$\frac{a_{\eta \eta}}{a}$ for tensor perturbations}\end{array}\right.
\end{equation}
Here $Z=a \dot{\phi}/H$ and $\eta$ is the  conformal time  defined, up to a constant, by  $dt=a\, d\eta$.

We now apply the psi series depending on $\varphi$ obtained above and several of their consequences to determine approximate expressions for the potentials $W_{\alpha}(\eta)$.

\subsubsection{Polynomial-exponential potentials}
For models with a  potential $v(\varphi)$ of the
form~(\ref{pot10})  we consider the  two-term approximation
\begin{equation}\label{pot0111}
h(\varphi)\approx \,e^{\varphi}\Big( \frac{1}{b}+ h_2(\varphi)e^{-2\varphi}\Big),
\end{equation}
of the psi series  (\ref{hpot1002}). The polynomial coefficient $h_2(\varphi)$ satisfies
\begin{equation}\label{achee}
h'_2-2\,h_2=-\frac{b}{2}\,v_0,
\end{equation}
so that it is given by
\begin{equation}\label{ache}
h_2(\varphi)=\frac{b}{2}\, e^{2\varphi}\int_{\varphi}^{+\infty} v_0(x)\, e^{-2x} d x.
\end{equation}
We show below    several psi series of the form
\begin{equation}\label{pot0111b}\everymath{\displaystyle}
f(\varphi)\,=\,e^{\alpha \varphi}\Big( \beta+\sum_{n=2}^{\infty}f_n(\varphi)e^{-n\varphi}\Big),
\end{equation}
with polynomial coefficients $f_n(\varphi)$,  which  are consequences of  (\ref{hpot1002}) and are required for our calculations .
\begin{enumerate}
\item The scale factor admits the expansion
\begin{equation}\label{pot0111a}\everymath{\displaystyle}
a(\varphi)\,=c\,\, e^{-\frac{\varphi}{3}}\Big( 1+\sum_{n=2}^{\infty}a_n(\varphi)e^{-n\varphi}\Big),
\end{equation}
where $c$ is an arbitrary nonzero positive constant and
\begin{equation}\label{a2}
a_2(\varphi)=\frac{b}{3}\,h_2(\varphi).
\end{equation}
\item The conformal time $\eta$, considered  as a function of $\varphi$, satisfies the differential equation
\begin{equation}\label{cont}
\frac{d \eta}{d \varphi}=-\frac{1}{a\,h'},
\end{equation}
and then it can be expanded as a psi series
\begin{equation}\label{pot0111eta}\everymath{\displaystyle}
 \eta(\varphi)\,=\, \frac{e^{-\frac{ 2\varphi}{3}}}{c}\Big( \frac{3}{2}\,b+\sum_{n=2}^{\infty}\eta_n(\varphi)e^{-n\varphi}\Big),
\end{equation}
with
\begin{equation}\label{eta}
\eta_2(\varphi)=b^2\, e^{\frac{8}{3}\varphi} \int_{\infty}^\varphi
\Big(h'_2(x)-\frac{2}{3}h_2(x)\Big) e^{-\frac{8}{3}x} d x.
\end{equation}
\item The Hubble radius  $1/aH=3/ah$   can be
expanded as
\begin{equation}\label{hra}
\frac{1}{aH}=\frac{3 b}{c}\, e^{-\frac{2}{3}\varphi}\Big( 1+\sum_{n=2}^{\infty}r_n(\varphi)e^{-n\varphi}\Big),
\end{equation}
where
\begin{equation}\label{ha2}
r_2(\varphi)=-\frac{4b}{3}\, h_2(\varphi).
\end{equation}
\item The function  $Z=a \dot{\phi}/H$ is proportional to
$z(\varphi)\equiv -a\,h'/h$, which has the psi series expansion
\begin{equation}\label{pot0111zeta}\everymath{\displaystyle}
z(\varphi)\,=\, -c\,\,e^{-\frac{ \varphi}{3}}\Big( 1+\sum_{n=2}^{\infty}z_n(\varphi)e^{-n\varphi}\Big),
\end{equation}
where
\begin{equation}\label{zeta2}
z_2(\varphi)=b\Big(h'_2(\varphi)-\frac{5}{3}h_2(\varphi)\Big).
\end{equation}
\end{enumerate}

Let us now consider the potentials $W_{\alpha}(\eta)$ (\ref{w2}). From  (\ref{pot0111eta}) we get the following approximation for
$\varphi(\eta)$
\begin{equation}\label{appph}
\varphi\approx -\frac{3}{2} \log \Big( \frac{2c}{3 b}\eta \Big).
\end{equation}
Inserting this result in (\ref{pot0111eta}) we get an  approximation with two terms for $\varphi(\eta)$ given by
\begin{equation}\label{appph1}
\varphi\approx -\frac{3}{2} \log \Big( \frac{2c}{3 b}\eta \Big)+\frac{1}{b}\Big( \frac{2c}{3 b}\eta \Big)^3 \eta_2 \Big( -\frac{3}{2} \log \Big( \frac{2c}{3 b}\eta \Big)  \Big).
\end{equation}
Then, using (\ref{pot0111zeta}) we obtain the two-term approximation
\begin{equation}\label{zapp}
z(\eta) \approx -c\Big(\frac{2c}{3b} \eta\Big)^{1/2}\Big(1+\tilde{z}_2(\log\eta) \eta^3 \Big), \quad \eta\rightarrow 0,
\end{equation}
where
\begin{equation}\label{zapp2}
\tilde{z}_2(\log\eta) \equiv \Big(\frac{2c}{3b}\Big)^3 \Big[ z_2 \Big(-\frac{3}{2} \log \Big( \frac{2c}{3 b}\eta \Big) \Big)
 -\frac{1}{3b}\eta_2 \Big(-\frac{3}{2} \log \Big( \frac{2c}{3 b}\eta \Big) \Big)\Big].
\end{equation}
Thus, we get that
\begin{equation}\label{apprR}
W_R(\eta)=\frac{z_{\eta \eta}}{z} \approx \frac{1}{\eta^2}\Big
[-\frac{1}{4}+\eta^3 \Big(\tilde{z}_2''(\log
\eta)+6\,\tilde{z}_2'(\log \eta)+9\tilde{z}_2(\log \eta) \Big)\Big
], \quad \eta\rightarrow
0.
\end{equation}

A completely similar calculation starting from (\ref{pot0111a}) and (\ref{a2}) leads to
\begin{equation}\label{apprT}
W_T(\eta)\approx \frac{1}{\eta^2}\Big [-\frac{1}{4}+\eta^3
\Big(\tilde{a}_2''(\log \eta)+6\,\tilde{a}_2'(\log
\eta)+9\tilde{a}_2(\log \eta) \Big)\Big ], \quad \eta\rightarrow
0,
\end{equation}
with
\begin{equation}\label{aapp2}
\tilde{a}_2(\log\eta)  \equiv \Big(\frac{2c}{3b}\Big)^3 \Big[ a_2 \Big(-\frac{3}{2} \log \Big( \frac{2c}{3 b}\eta \Big) \Big)
 -\frac{1}{3b}\eta_2 \Big(-\frac{3}{2} \log \Big( \frac{2c}{3 b}\eta \Big) \Big)\Big].
\end{equation}
For example, for the quadratic model~(\ref{quad}), from~(\ref{h1quad}),~(\ref{a2}),~(\ref{eta}),~(\ref{zeta2}) and~(\ref{zapp2})-(\ref{aapp2}) we obtain
\begin{equation}\label{WRTq}\everymath{\displaystyle}\begin{array}{rcl}
W_R(\eta)&\approx&-\frac{1}{4\eta^2}-\frac{\m^2c^3}{198b}\eta\left[131+756\log\left(\frac{2c}{3b}\eta\right)            +504\log\left(\frac{2c}{3b}\eta\right)^2\right],\\  \\
W_T(\eta)&\approx&-\frac{1}{4\eta^2}+\frac{\m^2c^3}{64b}\eta\left[-1+4\log\left(\frac{2c}{3b}\eta\right)
            +24\log\left(\frac{2c}{3b}\eta\right)^2\right].
\end{array}\end{equation}
\begin{figure}\label{figWq}
\begin{center}
        \includegraphics[width=7cm]{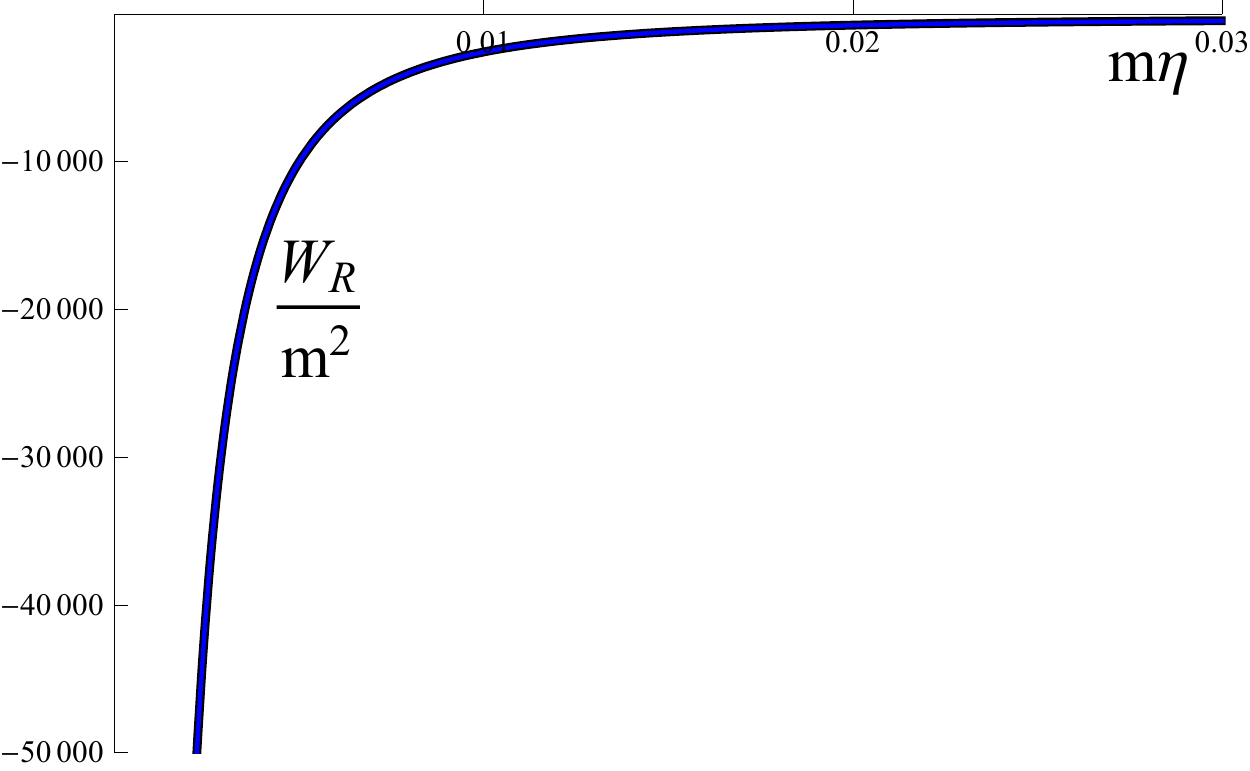}
        \hspace{1cm}
        \includegraphics[width=7cm]{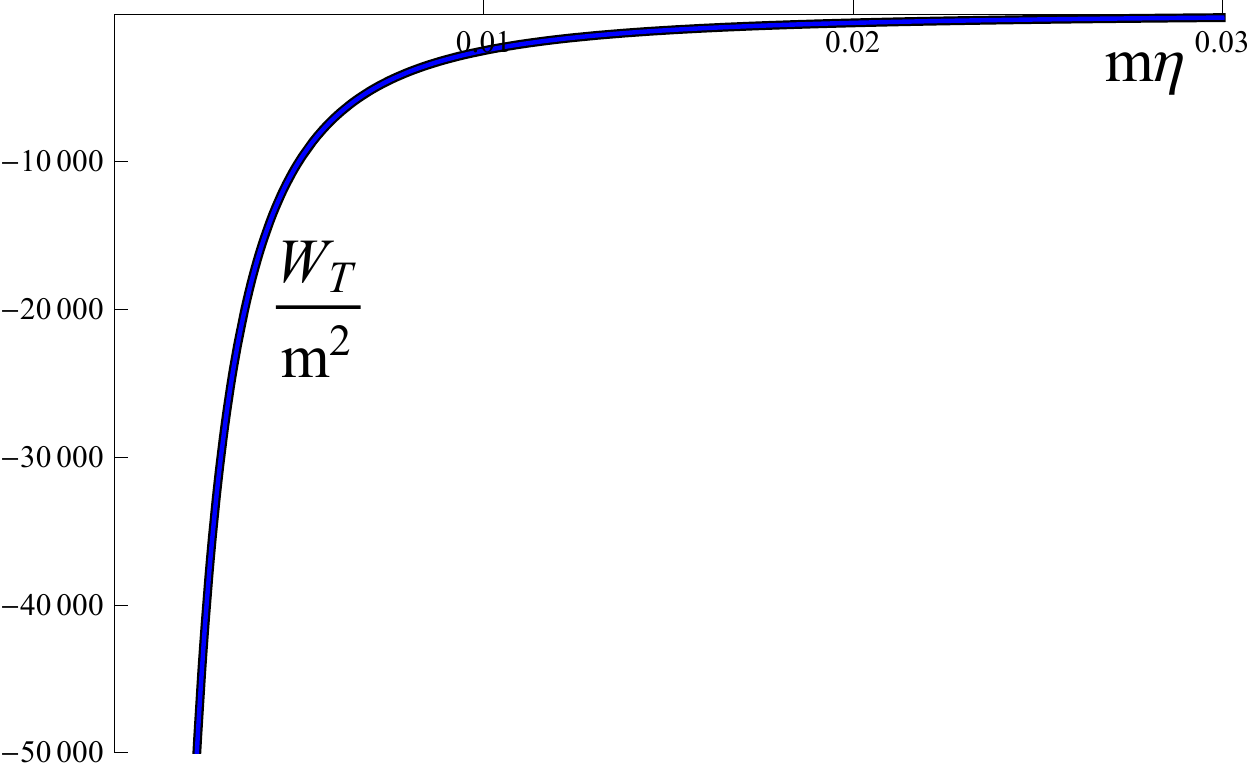}
\end{center}
\caption{Potentials $W_R(\eta)$ (left) and $W_T(\eta)$ (right) for
the quadratic model $v(\varphi)=\m^2\varphi^2$. The
black lines show the numerical approximations while the blue lines
show the psi series approximation~(\ref{WRTq}). Here
we have used the rescaled variables $\m\eta$, $\m^{-2}W_R$, $\m^{-2}W_T$, the value $b=b_{60}=\hat{b}_{60}/\m$,
$\hat{b}_{60}\approx 1.79104\times10^{7}$ obtained  in section~\ref{appA} for the quadratic
model and $c$ is such that $a(\varphi_{in})=1$.  We have taken a value  $\eta(t^*)=0$ of the conformal time   in agreement with~(\ref{pot0111eta}).}
\end{figure}

\subsubsection{Starobinsky  potentials}
For models with a potential $v(\varphi)$ of the form~(\ref{pot20})
we consider the  psi series  (\ref{pot20hub}) for $h(\varphi)$
and its truncation to two terms
\begin{equation}\label{pot20hubt}
h(\varphi)\,\approx
\,\frac{e^{\varphi}}{b}\Big(1+b^{2}\gamma_1(e^{-\alpha\varphi})e^{-2\varphi}\Big),
\end{equation}
where $\gamma_1$ is given by~(\ref{is1}) for $\alpha>0$ and by
setting the lower integration index as $\infty$ for
$\alpha\,\in\,(-\frac{2}{N},0)$. From  (\ref{pot20hub}) we
derive the psi series which follow.

\begin{enumerate}

\item The scale factor
\begin{equation}\label{pot20sf1}
a(\varphi)\,=\,c\,e^{-\frac{\varphi}{3}}\left(1+\sum_{n=1}^{\infty}b^{2n}a_n(u)e^{-2n\varphi}\right),
\end{equation}
with $a_n(u)$ being polynomials in $u$. In particular,
\begin{equation}\label{pot20a11}
a_1(u)\,=\,\frac{1}{3}\gamma_1(u).
\end{equation}
\item The conformal time
\begin{equation}\label{pot20ct}
\eta(\varphi)\,=\,\frac{3b}{2c}e^{-\frac{2\varphi}{3}}\left(1+\sum_{n=1}^{\infty}b^{2n}\eta_n(u)e^{-2n\varphi}\right),
\end{equation}
where $\eta_n(u)$ are polynomials in $u$. For example we have that
\begin{equation}\everymath{\displaystyle}\label{pot20eta1}
\eta_1(u)\,=\,\left\{\begin{array}{lll}
\frac{2}{3}\int_0^u\left(\gamma_1'(\mu)+\frac{2}{3\alpha\mu}\gamma_1(\mu)\right)
\Big(\frac{\mu}{u}\Big)^{\frac{8}{3\alpha}}d\mu, & & \alpha>0,\\  \\
-\frac{2}{3}\int_u^{\infty}\left(\gamma_1'(\mu)+\frac{2}{3\alpha\mu}\gamma_1(\mu)\right)
\Big(\frac{\mu}{u}\Big)^{\frac{8}{3\alpha}}d\mu, & &
\alpha\in(-\frac{2}{N},0).
\end{array}\right.\end{equation}
\item The Hubble radius
\begin{equation}\label{hrapot2}
\frac{1}{aH}\,=\,\frac{3b}{c}e^{-\frac{2\varphi}{3}}\left(1+\sum_{n=1}^{\infty}b^{2n}r_n(u)e^{-2n\varphi}\right),
\end{equation}
where $r_n(u)$ are polynomials in $u$. For instance
\begin{equation}\label{r1pot2}
r_1(u)\,=\,-\frac{4}{3}\gamma_1(u).
\end{equation}
\item The variable $z(\varphi)\equiv -a\,h'/h$
\begin{equation}\label{pot20muk}
z(\varphi)\,=\,-c\,e^{-\frac{\varphi}{3}}\left(1+\sum_{n=1}^{\infty}b^{2n}z_n(u)e^{-2n\varphi}\right),
\end{equation}
with $z_n(u)$ being polynomials in $u$. In particular
\begin{equation}\label{pot20z11}
z_1(u)\,=\,-\left(\alpha u \gamma_1'(u)+\frac{5}{3}\gamma_1(u)\right).
\end{equation}
\end{enumerate}
From~(\ref{pot20ct}) we can derive the  approximation
\begin{equation}\label{varphietapot20}
\varphi\,\approx\,-\frac{3}{2}\log\left(\frac{2c\eta}{3b}\right)+\frac{3b^2}{2}\left(\frac{2c\eta}{3b}\right)^3
\eta_1\left(\left(\frac{2c\eta}{3b}\right)^{\frac{3\alpha}{2}}\right).
\end{equation}
By replacing~(\ref{varphietapot20}) into~(\ref{pot20muk}), then  it is found that
\begin{equation}\label{zetaetapot20}
z(\eta)\,\approx\,-c\left(\frac{2c}{3b}\right)^{\frac{1}{2}}\eta^{\frac{1}{2}}\left[1+\eta^3\tilde{z}_1(\eta^{\alpha})\right],
\end{equation}
where
\begin{equation}\label{pot20z12}
\tilde{z}_1(\eta^{\alpha})\,=\,\frac{8c^3}{27b}\left[z_1\left(\left(\frac{2c\eta}{3b}\right)^{\frac{3\alpha}{2}}\right)-
\frac{1}{2}\eta_1\left(\left(\frac{2c\eta}{3b}\right)^{\frac{3\alpha}{2}}\right)\right].
\end{equation}
Similarly, from~(\ref{pot20sf1})-(\ref{pot20a11}) and~(\ref{varphietapot20}) we have that
\begin{equation}\label{pot20sf2}
a(\eta)\,\approx\,c\left(\frac{2c}{3b}\right)^{\frac{1}{2}}\eta^{\frac{1}{2}}\left[1+\eta^3\tilde{a}_1(\eta^{\alpha})\right],
\end{equation}
with
\begin{equation}\label{pot20a12}
\tilde{a}_1(\eta^{\alpha})\,=\,\frac{8c^3}{27b}\left[\frac{1}{3}\gamma_1\left(\left(\frac{2c\eta}{3b}\right)^{\frac{3 \alpha}{2}}\right)-
\frac{1}{2}\eta_1\left(\left(\frac{2c\eta}{3b}\right)^{\frac{3 \alpha}{2}}\right)\right].
\end{equation}
Thus,~(\ref{zetaetapot20}) leads to
\begin{equation}\label{WRpot20}
W_R(\eta)\approx\,\frac{1}{\eta^2}\left[-\frac{1}{4}+\eta^3\left(\alpha^2\eta^{2\alpha}\tilde{z}_1''(\eta^\alpha)
+\alpha(\alpha+6)\eta^{\alpha}\tilde{z}_1'(\eta^{\alpha})+9\tilde{z}_1(\eta^{\alpha})\right)\right], \quad \eta\rightarrow
0.
\end{equation}
Analogously, from~(\ref{pot20sf2}) one gets
\begin{equation}\label{WTpot20}
W_T(\eta)\approx\,\frac{1}{\eta^2}\left[-\frac{1}{4}+\eta^3\left(\alpha^2\eta^{2\alpha}\tilde{a}_1''(\eta^{\alpha})
+\alpha(\alpha+6)\eta^{\alpha}\tilde{a}_1'(\eta^{\alpha})+9\tilde{a}_1(\eta^{\alpha})\right)\right],\quad \eta\rightarrow
0
.\end{equation}
For example, for the Starobinsky model~(\ref{sta}), from the first equation~(\ref{stagammas}),~(\ref{pot20a11}),~(\ref{pot20eta1}),~(\ref{pot20z11}),~(\ref{pot20z12}) and~(\ref{pot20a12})-(\ref{WTpot20}), it is found that
\begin{equation}\label{WRTS}\everymath{\displaystyle}\begin{array}{rcl}
W_R(\eta)&\approx&-\frac{1}{4\eta^2}-\frac{\lambda c^3}{3b}\eta\Big[\frac{2(\alpha+1)(6\alpha^2+14\alpha+7)}{3\alpha+4}
           \left(\frac{2c\eta}{3b}\right)^{3\alpha}\\  \\
         & &\qquad -\frac{2(\alpha+2)(3\alpha^2+14\alpha+14)}{3\alpha+8}\left(\frac{2c\eta}{3b}\right)^{\frac{3\alpha}{2}}+
                   \frac{7}{2}\Big],\\ \\
W_T(\eta)&\approx&-\frac{1}{4\eta^2}+\frac{\lambda c^3}{3b}\eta\Big[\frac{2(\alpha+1)}{3\alpha+4}\left(\frac{2c\eta}{3b}\right)^{3\alpha}-\frac{4(\alpha+2)}{3\alpha+8}\left(\frac{2c\eta}{3b}\right)^{\frac{3\alpha}{2}}
+\frac{1}{2}\Big].
\end{array}\end{equation}

We notice that, as it was found in ~\cite{DE10},  the common dominant term near the singularity $\eta=0$ of the potentials (\ref{apprR}), (\ref{apprT}),
(\ref{WRpot20}) and (\ref{WTpot20})
 coincides with  the critical central singular attractive potential allowing the fall to the center of a quantum particle.

 Finally, we point out that we may also use our psi-series in the variable $\varphi$ starting from the Mukhanov-Sasaki equation  (\ref{perct}) with the potential $W_{\alpha}$ expressed as a function of $\varphi$.  However,  as it is showed in Equation (2.27) of  ~\cite{DE10}, the function  $W_{\alpha}(\varphi)$ is rather involved, so that  we do not find  any advantage in using our psi-series in that way.
\begin{figure}\label{figWS}
\begin{center}
        \includegraphics[width=7cm]{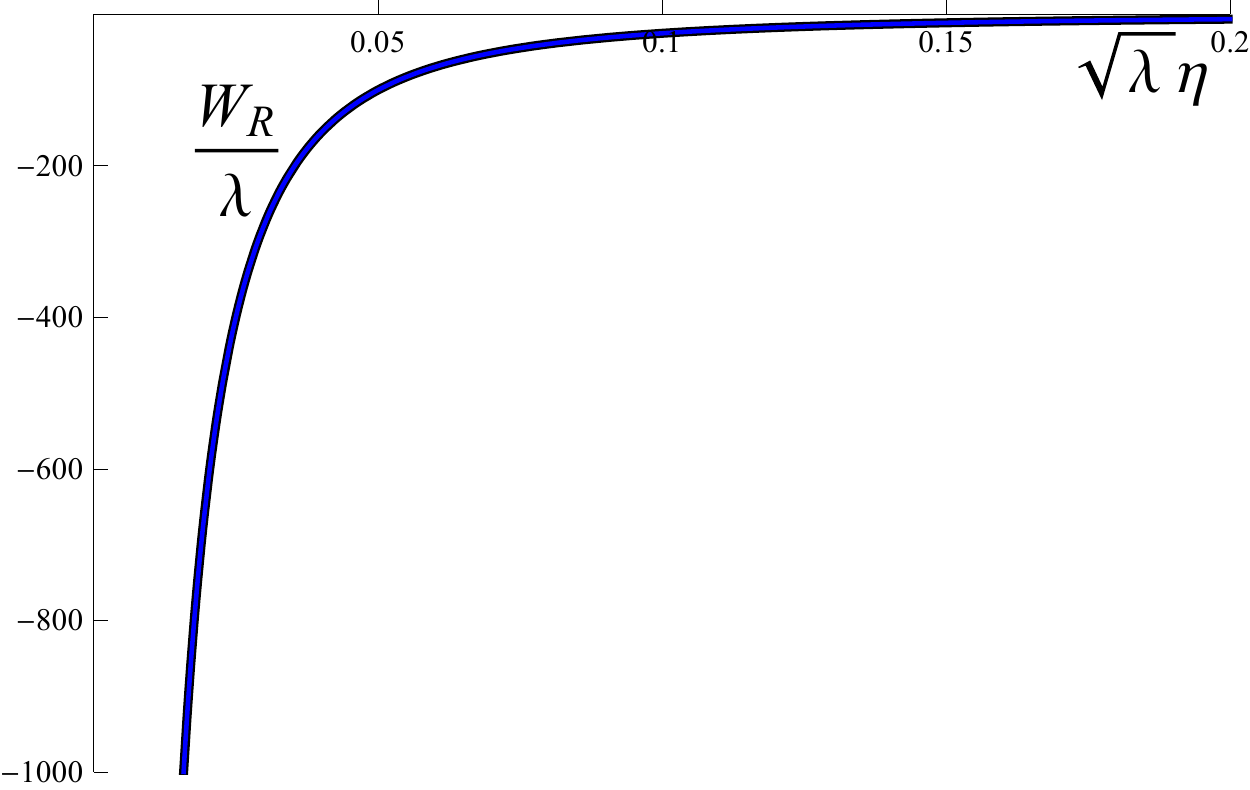}
        \hspace{1cm}
        \includegraphics[width=7cm]{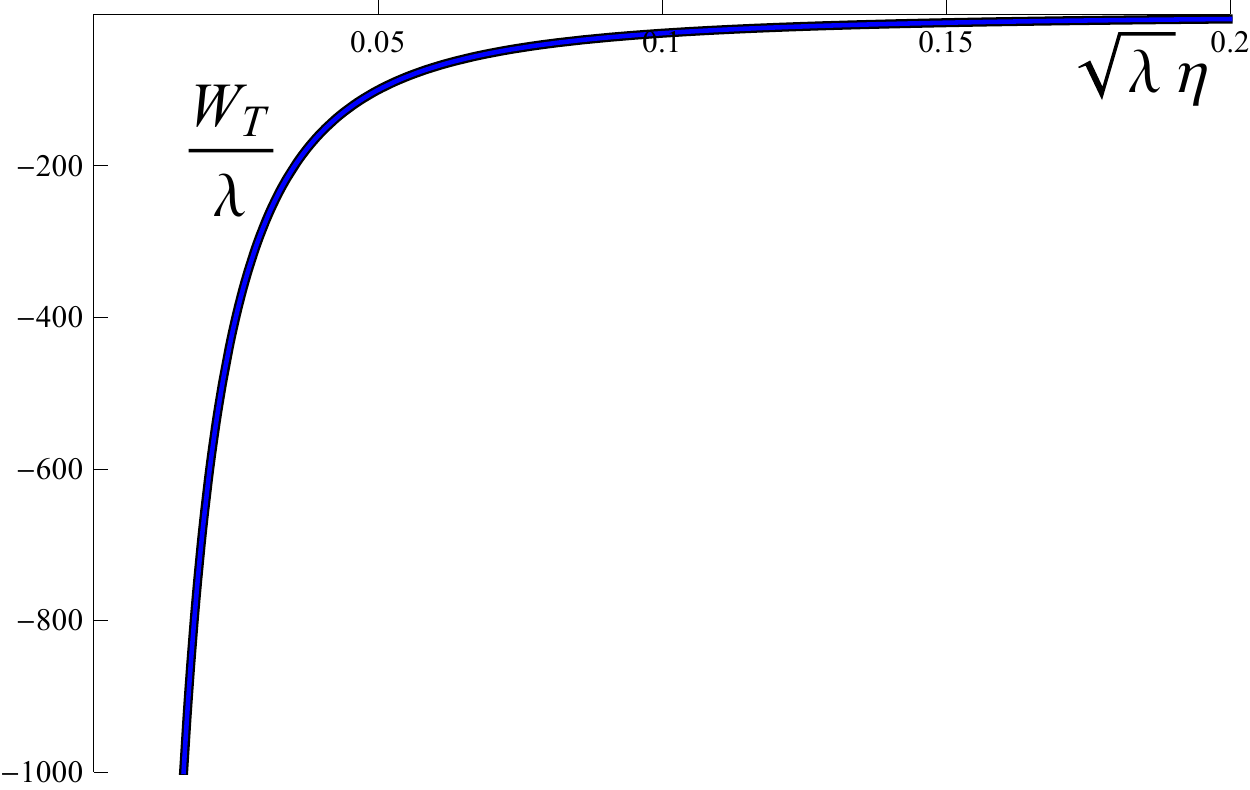}
\end{center}
\caption{Potentials $W_R(\eta)$ (left) and $W_T(\eta)$ (right) for
the Starobinsky potential
$v(\varphi)=\lambda(1-e^{-\frac{\varphi}{\sqrt{3}}})^2$. The black
lines show the numerical approximations while the blue lines show
the psi series approximation~(\ref{WRTS}). Here we have used the rescaled variables $\sqrt{\lambda}\eta$, $\lambda^{-1}W_R$,
$\lambda^{-1}W_T$ and the value
 $b=\hat{b}_{60}/\sqrt{\lambda}$, $\hat{b}_{60}\approx 2649.03$ obtained in section~\ref{appA},
 $c$ is taken so that $a(\varphi_{in})=1$ and  $\eta(t^*)=0$ in agreement with~(\ref{pot20ct}).}
\end{figure}

\section{CONCLUSIONS}\label{con}

In  this paper, we have developed  a method to determine psi-series formal  solutions  of  single field inflation
models~(\ref{eq1})-(\ref{eq2}) during  the kinetic dominance period. The scheme is based on  the Hamilton-Jacobi formalism of  inflaton
models~\cite{SA90, LI00, SEP} and  provides psi-series  depending on the inflaton field.
The method has been applied to models with polynomial-exponential potential functions~(\ref{pot10}) (two particularly important examples are the quadratic
potential~(\ref{quad}) and the Higgs potential~(\ref{Higgs})) and to models with generalized Starobinsky potential
functions~(\ref{pot20}) (including the standard Starobinsky potential~(\ref{sta})). We have also proved that  there exist psi series  near the singularity  for several  physically relevant quantities such  as the scale factor, the conformal time and the Hubble radius. The explicit form of the first two terms of these expansions  has been given.

 We have found that truncations of these psi series can be used  to determine the value of the inflaton field at the initial moment of the inflation period, and to include the effect of the KD period to estimate  the amount of inflation. Furthermore, we have shown that psi series can be applied to determine explicit corrections  depending on the inflaton field to the dominant term of the potentials of the Mukanov-Sasaki equation for both, curvature and tensor perturbations.

\vspace{0.3cm}
\noindent
{\bf ACKNOWLEDGEMENTS}

The financial support of the Spanish Ministerio de Econom\'{i}a y Competitividad under Projects No. FIS2015-63966-P and PGC2018-094898-B-I00 is gratefully acknowledged.  We thank Professor Gabriel Alvarez Galindo for fruitful discussions.

\bibliography{inflation-gpsi}

\end{document}